\documentclass[english]{article}
\usepackage[T1]{fontenc}
\usepackage[latin9]{inputenc}
\usepackage{babel}
\usepackage{amsmath, amsthm, amssymb, mathrsfs}
\usepackage{dsfont}
\usepackage{thmtools, thm-restate}
\usepackage{enumerate}
\usepackage{footmisc}
\usepackage{color}
\usepackage{fancyhdr}
\usepackage{bussproofs}
\usepackage{graphicx}
\usepackage[margin=1in]{geometry}
\usepackage{caption}
\usepackage[natbibapa]{apacite}
\usepackage{hyperref}
\usepackage{placeins}
\usepackage{setspace}
\usepackage{tikz}
\usepackage{environ}
\makeatletter

\newtheorem{theorem}{Theorem}[section]
\newtheorem{lemma}[theorem]{Lemma}
\newtheorem{proposition}[theorem]{Proposition}

\theoremstyle{definition}

\newtheorem{definition}[theorem]{Definition}

\theoremstyle{remark}
\newtheorem{remark}[subsection]{Remark}

\usepackage{eurosym}
\usepackage{amssymb}

\usepackage{verbatim}
\usepackage{amsmath}
\usepackage{tikz}
\usepackage{mathrsfs}
\usepackage[margin=10pt,font=small,labelfont=bf]{caption}

\usepackage{graphicx}
\usepackage{dsfont}
\usepackage{enumerate}

\usepackage[title,toc,titletoc,header]{appendix}

\renewcommand{\phi}{\varphi}

\usepackage{tikz}
\usetikzlibrary{calc,matrix,positioning,fit, shapes, automata, arrows, patterns}

\tikzset{
  solid node/.style={circle,draw,inner sep=1.2,fill=black},
  empty node/.style={font=\tiny, circle, outer sep=0pt, inner sep=0pt},
  matrix node/.style={rectangle, inner sep=0pt, draw, minimum height=0.5cm, minimum width=0.8cm, outer sep=0pt},  
  noline/.style={edge from parent/.style={draw=none}},
line/.style={edge from parent/.style={draw}},
}

\title{From \emph{ATL} to Stit Theory}
\date{}
\author{Aldo Iv\'an Ram\'irez Abarca}
\begin{document}
\maketitle

\section{\emph{ATL} Syntax and Semantics}

Developed by Alur et al. throughout a series of papers  \citep[see, for instance,][]{alur1997alternating, AlurHenzingerKupferman2002}), \emph{alternating-time temporal logic} was presented as an extension of the so-called $\mathtt{CTL}$, a branching-time temporal logic with two modal operators quantifying over sets of \emph{paths}---sequences of states such that each element of the sequence transitions into the next---such that formulas involving universal and existential quantification of these paths were used to reason about properties of computations in a system. Based on these ideas, alternating-time temporal logic was introduced to reason about \emph{strategies} over such transition systems, where the main question involved formalizing when a coalition of agents is able to choose and perform a strategy---where the term strategy is used to refer to a set of alternative sequences of transitions, alternative according to the path they involve---such that $\phi$ is guaranteed to occur on specific states of all the paths (or computations) belonging to the strategy. The formal definitions for this logic are included below. 

  \begin{definition}[Syntax of $ATL$]
\label{syntaxatl}
Given a finite set $Ags$ of agent names and a countable set of propositions $P$, the grammar for the formal language $\mathcal L_{\textsf{ATL}}$ is given by
\[ \begin{array}{lcl}
\phi :=  p \mid \neg \phi \mid \phi \wedge \psi \mid \langle\langle C\rangle\rangle X\phi \mid \langle\langle C\rangle\rangle \mathsf{G}\phi \mid \langle\langle C\rangle\rangle \phi\mathsf{U}\psi,
\end{array} \]
where $p \in P$ and $C\subseteq Ags$.
\end{definition}

 $[C] \varphi$---where $\phi$ is a temporal-logic formula of the form $X\psi$, $\mathsf{G}\psi$, or $\psi\mathsf{U}\theta$---is meant to express  that coalition $C$ has the ability to ensure that $\phi$ is the case, regardless of what the agents in $Ags-C$ can do. In this setting,  $X\psi$ is meant to express that $\psi$ holds at the next state, $\mathsf{G}\psi$ is meant to express that $\psi$ holds henceforward (at the present state and at all the states along a given computation), and $\psi\mathsf{U}\theta$ is meant to express that $\theta$ will hold at some future state along a given computation, and until this happens, such a computation will involve states where $\phi$ holds. 
 
 As for the semantics, the formulas of $\mathcal{L}_{\textsf{ATL}}$ are typically evaluated using either \emph{alternating-time structures} (ATS's) or \emph{concurrent game structures} (CGS's) ATS's and CGS's were shown to be equivalent by \citet{goranko2004comparing}, who showed that the logic  is the same over both classes of structures. Here, I will present $ATL$ using CGS's.    
 
 \begin{definition}[Concurrent game structures (CGS's)]
 \label{cgs}
 A tuple $\mathcal{S}:=\langle W, Ags, Act, \delta,\mathcal{V}\rangle$ is a \emph{concurrent game structure} (CGS) iff \begin{itemize}
     \item $W$ is a finite set of \emph{states}.
     
     The set of all infinite sequences of elements in $W$ will be denoted $W^{\infty}$. The elements of this set are known as \emph{paths} or \emph{computations}. The set of all finite sequences of elements in $W$ is denoted by $W^+$. The elements of $W^+$ are finite prefixes of computations, which will be denoted in the form $\lambda=w_1,\dots, w_k$ for some $k\in \mathds{N}$. 
     
     For a given $\lambda \in W^+$, I will denote the length of $\lambda$ by $len(\lambda)$. For $1\leq i \leq len(\lambda)$, $\lambda(i)$ denotes the $i$\textsuperscript{th} element in the sequence $\lambda$, and $\lambda[1,i]$ denotes the initial segment of $\lambda$ up to the $i$\textsuperscript{th} element.

     \item $Act$ is a function that assigns to each agent $\alpha\in Ags$ and state $w\in W$ a finite non-empty set $Act_\alpha^w$ of \emph{actions labels}, of the form $\{s_{\alpha,w}, s_{\alpha,w}',s_{\alpha,w}'',\dots\}$. $Act_\alpha^w$ is interpreted as the set of actions available to $\alpha$ at $w$. 
     
     For coalition $C\subseteq Ags$ and $w\in W$, the set  $Act_C^w:=\Pi_{\alpha\in C} Act_\alpha^w$ is interpreted as the set of joint actions available to $C$ at $w$. The elements of this set will be denoted in the form $\mathbf{s}_{C,w}=\langle s_{\alpha,w} \rangle_{\alpha\in C}$, and $\mathbf{s}_\alpha$ will abbreviate $\mathbf{s}_{\{\alpha\}}$.  The set $Act_{Ags}:=\bigcup_{w\in W}Act_{Ags}^w$ is referred to as the set of \emph{action profiles over $\mathcal{S}$}. For $\mathbf{s}\in Act_{Ags}$, the projection of $\mathbf{s}$ along index $\alpha$ is denoted by $(\mathbf{s})_\alpha$.
     
     \item $\delta:W\times Act_{Ags}\to W$ is a function, known as the \emph{transition function}, mapping each state and action profile to a unique state. For  $w\in W$ and  joint action $\mathbf{s}_{C,w}$ (where $C\subseteq Ags$), the set $\left\{v\in W; v=\delta(w,\mathbf{s}_{Ags,w})  \mbox{ for some } \mathbf{s}_{Ags,w}\in Act_{Ags}^w \mbox{ s. t. } (\mathbf{s}_{Ags,w})_\alpha=(\mathbf{s}_{C,w})_\alpha \mbox{ for every } \alpha\in C\right\}$ will be denoted by $\delta[w, \mathbf{s}_{C,w}]$, and will be referred to as the set of the possible \emph{successor states} of joint action $\mathbf{s}_{C,w}$ at $w$. 
     
     Observe that, for each $w\in W$ and action profile $\mathbf{s}_{Ags,w}$, $\delta[w, \mathbf{s}_{Ags,w}]$ is a singleton, which will be denoted by $w^{+\mathbf{s}_{Ags,w}}$.

     \item $\mathcal{V}:P\to 2^W$ is a valuation function.
      \end{itemize}
      \end{definition}
     
     In order to provide the semantics for the formulas of $\mathcal{L}_{\textsf{ATL}}$ on CGS'S, further definitions are required. 
    
    \begin{definition}[Strategies in $ATL$]
     For a given agent $\alpha$, a \emph{strategy of $\alpha$ over $\mathcal{S}$} is a function $f_\alpha:W^+\to \bigcup_{w\in W}Act_\alpha^w$ that assigns to each finite sequence of the form $\lambda=w_1,\dots,w_k$ an element in $Act_\alpha^{w_k}$. The set of all strategies of $\alpha$ over $\mathcal{S}$ will be denoted by $Strat_\alpha$. For coalition $C\subseteq Ags$, a \emph{collective strategy of $C$ over $\mathcal{S}$} is defined as a tuple of the form $f_C:=\langle f_\alpha \rangle_{\alpha\in C}$ (where $f_\alpha$ is a strategy for $\alpha$ over $\mathcal{S}$ for every $\alpha\in C$). The set of all strategies of $C$ over $\mathcal{S}$ will be denoted by $Strat_C$. 
     
     For $\lambda\in W^+$ and  collective strategy $f_C$, the joint action given by $\langle f_\alpha(\lambda) \rangle_{\alpha\in C}$ will be denoted by $f_C(\lambda)$.  The \emph{outcome} from $w$ of a collective strategy $f_C$---$out(w, f_C)$---is defined as follows:\[out(w,f_C):=\left\{\mu\in W^{\infty}; \begin{array}{l}
           \mu(1)=w,\\
           \mu(i+1)\in\delta[\mu(i),f_C(\mu[1,i])]\mbox{ for every } i\in \mathds{N}-\{0\}
        \end{array} \right\}.\] 
        
        For $w\in W$, $f_C\in Strat_{C}$, and $\mu\in out(w, f_C)$, I will denote by $\mu^{fin}$ the set $\left\{\mu[1,i]; i\in \mathds{N}-\{0\}\right\}$. Observe that, for each $f_{Ags}\in Strat_{Ags}$, $out(w,f_{Ags})$ is a singleton. 
     \end{definition}
   
\begin{definition}[Evaluation rules for $ATL$]
    
 Let $\mathcal{S}$ be a CGS. The semantics on $\mathcal{S}$   for the formulas of $\mathcal {L}_{\textsf{ATL}}$ are defined recursively by the following truth conditions, evaluated at state $w$:
\[ \begin{array}{lll}
\mathcal{S},w \models p & \mbox{iff} & w \in \mathcal{V}(p) \\

\mathcal{S},w \models \neg \phi & \mbox{iff} & \mathcal{S},w \not\models \phi \\

\mathcal{S},w\models \phi \wedge \psi & \mbox{iff} & \mathcal{S},w \models \phi \mbox{ and } \mathcal{S},w \models \psi \\

\mathcal{S},w \models \langle\langle C\rangle\rangle X\phi &
\mbox{iff} & \mbox{there exists } f_C \mbox{ s.t. } \mathcal{S},\lambda(2) \models \phi \\&& \mbox{for every } \lambda\in out(w,f_C) \\

\mathcal{S},w \models \langle\langle C\rangle\rangle \mathsf{G}\phi &
\mbox{iff} & \mbox{there exists } f_C \mbox{ s.t. } \mathcal{S},\lambda(i) \models \phi \\&& \mbox{for every }  \lambda\in out(w,f_C) \mbox{ and every } i\in \mathds{N}\\

\mathcal{S},w \models \langle\langle C\rangle\rangle \phi\mathsf{U}\psi &
\mbox{iff} & \mbox{there exists } f_C \mbox{ s.t., for each } \lambda\in out(w,f_C), \mbox{ there is } \\&& j\in \mathds{N}-\{0\}   \mbox{ s.t. } \mathcal{S},\lambda(j) \models \psi \mbox{ and }  \mathcal{S},\lambda(i) \models \phi \\&& \mbox{for every }1\leq j< i.\\

    \end{array} \]
 \end{definition}
 
 Therefore, CGS's are very similar to \emph{coalition logic}'s game models \citep{pauly2002modal}. In essence, CGS's are nothing more than game models for which paths---or computations---and coalitions' strategies are also defined. The quantification over paths and strategies, then, is what underlies the semantics of the formulas of $\mathcal{L}_{\textsf{ATL}}$. Just as in the case of coalition logic, the semantics for the formulas of $\mathcal{L}_{\textsf{ATL}}$ is standard for atomic propositions and for formulas built with the Boolean connectives. The novel modality is given by formulas of the form $\langle\langle C\rangle\rangle\phi$, where $\phi$ represents one of three modalities of linear temporal logic ($X\psi$, $\mathsf{G}\psi$, or $\psi\mathsf{U}\theta$). The semantics for $\langle\langle C\rangle\rangle\phi$ is explained as follows: at state $w$ coalition $C$ is able to ensure that $\phi$ holds---at those states implied by the modality $\phi$ of linear temporal logic---iff for every member of $C$ there exists a strategy such that each computation in the outcome set of the collective strategy given by the profile of those individual strategies satisfies $\psi$ ($\theta$), at said states.

 \section{Embedding \emph{ATL} into Stit Theory}

\citet{broersen2006embedding} explored a formal relation between $ATL$ and stit theory, and he gave the essential ideas in order to embed $ATL$ into stit theory. Since the details of such an embedding are missing in the literature, I decided to include them in here, thus settling once and for all certain questions concerning the connection between concurrent game structures, on the one hand, and labelled \emph{bdt}-models for \emph{group xstit theory with strategic ability}, on the other. Let us begin with the formal definitions.

\begin{definition}\label{synta}
Given a finite set $Ags$ of agent names and a countable set of propositions $P$ such that $p \in P$ and $C\subseteq Ags$, the grammar of the formal language $\mathcal{L}_{\mathtt{SX}}$ is given by:
\[ \begin{array}{lcl}
\phi :=  p \mid \neg \phi \mid \phi \wedge \psi \mid X\phi \mid \mathtt{G}\phi \mid \phi\mathtt{U}\psi \mid \square\phi \mid [C]\phi \mid  \langle\langle C\rangle\rangle^s\phi.
\end{array} \]

 $X\phi$ stands for `$\phi$ holds at the next moment (along the same history).' $\mathtt{G}\phi$ stands for `$\phi$ holds now and at all future moments (along the same history).' $\phi\mathtt{U}\psi$ stands for `$\phi$ holds until $\psi$ holds.'  $\Box\varphi$ is meant to express the `historical necessity' of $\varphi$
($\Diamond \varphi$ abbreviates $\neg \Box \neg \varphi$).  $[C] \varphi$ stands for `coalition $C$ sees to it that $\varphi$.' $\langle \langle C\rangle \rangle^s \varphi$ is meant to express that `coalition $C$ has the strategic ability to ensure that $\phi$ is the case, regardless of what the agents in $Ags-C$ can do.' 
\end{definition}

 As for the semantics, the formulas of $\mathcal{L}_{\mathtt{SX}}$ will be evaluated using structures that I refer to as \emph{labelled \emph{bdt}-models}:

\begin{definition}[Labelled \emph{bdt}-models]\label{labelbdt}

 $\mathcal{M}:=\langle M,\sqsubset, Ags, \mathbf{Choice}, Tps, Lbl, Exe,\mathcal{V}\rangle$ is called a labelled \emph{bdt}-model iff
\begin{itemize}
    \item $M$ is a non-empty set of \textnormal{moments} and $\sqsubset$ is a strict partial ordering on $M$ satisfying \emph{no backward branching}: for all $m, m', m''\in M$ such that $m'\sqsubset m$ and $m''\sqsubset m$, either $m'=m''$ or $m'\sqsubset m''$ or $m''\sqsubset m'$. Each maximal $\sqsubset$-chain is called a $\textnormal{history}$, and the set of all histories is denoted by $H$. For $m\in M$, $H_m:=\left\{h \in H ;m\in h\right\}$. Tuples $\left\langle m,h \right\rangle$ such that $m \in M$, $h \in H$, and $m\in h$, are called \emph{indices}, and the set of indices is denoted by $I(M\times H)$.

    These structures are called `discrete-time' because $(M,\sqsubset)$ must additionally meet the following requirement:\begin{itemize}
        \item $(\mathtt{TD})$ \emph{Time-discreteness}: for all $m\in M$ and $h\in H_m$, there exists a unique moment $m^{+h}$ such that $m\sqsubset m^{+h}$ and $m^{+h}\sqsubseteq m'$ for every $m'\in h$ such that $m\sqsubset m'$. For $m\in M$ and $h\in H_m$, $m^{+h}$ is known as the \emph{successor of $m$ along $h$}. For an index $\left\langle m, h\right\rangle$, I refer to $\left\langle m^{+h}, h \right\rangle$ as the \emph{successor of $\left\langle m, h\right\rangle$}. For $m\in M$ and $h\in H_m$, the moment $m^{+h}$ will also be denoted by $m^{+h(1)}$, the moment $\left(m^{+h}\right)^{+h}$ will also be denoted by $m^{+h(2)}$, so that for each $i\in \mathds{N}-\{0\}$, $m^{+h(i)}$ will denote the unique moment in $h$ that is the $i$\textsuperscript{th} iteration of the successor function applied to $m$. For the sake of coherence, $m^{+h(0)}$ will also denote $m$.
    \end{itemize}  

$Ags$ is the finite set of agent names from Definition~\ref{synta}.

\item $\mathbf{Choice}$ is a function that maps each agent $\alpha$ and moment $m$ to a  partition $\mathbf{Choice}^m_\alpha$ of $H_m$, where the cells of such a partition represent $\alpha$'s available choices of action at $m$. For $m\in M$ and $h\in H_m$, $\mathbf{Choice}_\alpha^m(h)$ denotes the cell that includes $h$. This cell represents the choice of action that $\alpha$ has performed at index $\left\langle m,h\right\rangle$, and I refer to it as $\alpha$'s \emph{current choice of action at $\left\langle m,h\right\rangle$}. $\mathbf{Choice}$ satisfies two conditions: \begin{itemize}
\item $(\mathtt{NC})$ \emph{No choice between undivided histories}: for all $\alpha\in Ags$ and $h, h'\in H_m$, if $m'\in h\cap h'$ for some $m' \sqsupset m$, then $h\in L$ iff $h'\in L$ for every $L\in \mathbf{Choice}^m_\alpha$.

\item $(\mathtt{IA})$ \emph{Independence of agency}: a function $s:Ags\to 2^{H_m}$ is called a \emph{selection function at $m$} if it assigns to each $\alpha$ a member of $\mathbf{Choice}^m_\alpha$. If $\mathbf{Select}^m$ denotes the set of all selection functions at $m$, then, for all $m\in M$ and $s\in\mathbf{Select}^m$, $\bigcap_{\alpha \in Ags} s(\alpha)\neq \emptyset$. 

This condition establishes that concurrent actions
by distinct agents must be independent: the choices of action of a given 
agent cannot affect the choices available to another \citep[see][for a discussion of this property]{belnap01facing,horty1995deliberative}.
\end{itemize}

 \item $Tps$ is a set of \emph{action types}. For $\alpha\in Ags$ and $m\in M$, $Tps_\alpha^m$ denotes the set of action types that are available to $\alpha$ at $m$. 
    \item $Lbl$ is a label function that maps action tokens to action types: for $\alpha\in Ags$, $m\in M$, and $L\in\mathbf{Choice}_\alpha^m$, $Lbl(L)\in Tps$. For $\alpha\in Ags$, $Lbl_\alpha$ will denote a function that maps an index to the action label of the action token performed by $\alpha$ at that index. In other words, for index $\langle m, h \rangle$, $Lbl_\alpha\left(\langle m, h \rangle\right)=Lbl\left(\mathbf{Choice}_\alpha^m\left(h\right)\right)$. 
    
    
    
    \item $Exe$ is a partial execution function that maps each action type $\tau\in Tps$, $m\in M$, and $\alpha\in Ags$ to a particular action token $Exe^m_\alpha(\tau)\in\mathbf{Choice}^m_\alpha$.

    $Lbl$ and $Exe$ satisfy the following conditions: 
    \begin{itemize}
        \item $(\mathtt{EL})$ For each $\alpha\in Ags$ and index $\left\langle m, h \right\rangle$,  $Exe^m_\alpha(Lbl_\alpha(\left\langle m, h \right\rangle))=\mathbf{Choice}^m_\alpha(h)$.
        \item $(\mathtt{LE})$ For each $\alpha\in Ags$, $m\in M$, and $\tau\in Tps$, if $Exe^m_\alpha(\tau)$ is defined, then $Lbl_\alpha(Exe^m_\alpha (\tau))=\tau.$
    \end{itemize}
    \item  $ \mathcal{V}: P\to   
    2^{M \times H}$ is a valuation function that assigns to each atomic proposition a set of indices.
        \end{itemize}
        Labelled \emph{bdt}-models are called \emph{deterministic} if they  satisfy the following condition: for each $m\in M$ and $h\in H_m$, if $h'\in \mathbf{Choice}_{Ags}^m(h)$, then $m^{+h}=m^{+h'}$.
\end{definition}

In order to present the semantics for the formulas of $\mathcal{L}_{\textsf{SX}}$, with respect to labelled \emph{bdt}-frames, further definitions are required. 

\begin{definition}[Group strategies in stit theory]
Let $\mathcal{M}$ be a labelled \emph{bdt}-model with set of moments $M$.

\begin{itemize}
\item For coalition $C\subseteq Ags$,  $m\in M$, and $h\in H_m$, the set $\bigcap_{\alpha \in C}  \mathbf{\mathbf{Choice}}_\alpha^m(h)$ will be denoted by $\mathbf{\mathbf{Choice}}_{C}^m(h)$, so that the set  $\mathbf{\mathbf{Choice}}_{C}^m:=\{\mathbf{\mathbf{Choice}}_{Ags}^m(h);h\in H_m\}$ is interpreted as the partition of actions available to coalition $C$ at $m$.

\item For $\alpha\in Ags$ and $m\in M$, a \emph{strategy of $\alpha$ starting at $m$} is defined as a function $\sigma: \sqsubseteq[m]\to \bigcup_{m'\in \sqsubseteq[m]}\mathbf{Choice}^{m'}_\alpha$, where $\sqsubseteq[m]:=\{m'\in M; m\sqsubseteq m'\}$. For each coalition $C\subseteq Ags$, a \emph{collective strategy of $C$  starting at $m$} is defined as a tuple of the form $\sigma_C:=\langle \sigma_\alpha \rangle_{\alpha\in C}$ (where $\sigma_\alpha$ is a strategy of $\alpha$, starting at $m$, for every $\alpha\in C$). I will write $\sigma_C(m')$ to refer to the intersection $\bigcap_{\alpha\in C}\left(\sigma_C\right)_\alpha(m')$, which is an element in $\mathbf{Choice}^{m'}_C$. 

\item For $C\subseteq Ags$, $m\in M$, and collective strategy $\sigma_C$ starting at $m$, the set \[\mathbf{Adm}_C^m\left(\sigma_C\right):=\left\{h'\in H_m; \begin{array}{l}

     h'\in \sigma_C\left(m^{+h'(i)}\right)
     \\\mbox{for every } i\in \mathds{N}
\end{array} \right\}\]

is known as the set of \emph{admissible} histories of $\sigma_C$. 
\end{itemize}

\end{definition}

Labelled \emph{bdt}-models allow us to provide semantics for the formulas of $\mathcal{L}_{\textsf{SX}}$:

\begin{definition}[Evaluation rules]
\label{labelledbdtmodels}
Let labelled \emph{bdt}-model $\mathcal {M}$. The semantics on $\mathcal {M}$ for the formulas of $\mathcal {L}_{\textsf{SX}}$ are defined recursively by the following truth conditions, evaluated at a given index $\langle m,h \rangle$:
\[ \begin{array}{lll}
\mathcal{M},\langle m,h \rangle \models p & \mbox{iff} & \langle m,h \rangle \in \mathcal{V}(p) \\

\mathcal{M},\langle m,h \rangle \models \neg \phi & \mbox{iff} & \mathcal{M},\langle m,h \rangle \not\models \phi \\

\mathcal{M},\langle m,h \rangle \models \phi \wedge \psi & \mbox{iff} & \mathcal{M},\langle m,h \rangle \models \phi \mbox{ and } \mathcal{M},\langle m,h \rangle \models \psi \\

\mathcal{M},\langle m,h \rangle \models  X \varphi &
\mbox{iff}& \mathcal{M},\langle m ^{+h},h \rangle \models \varphi \\

\mathcal{M},\langle m,h \rangle \models  \mathsf{G} \varphi &
\mbox{iff}& \mathcal{M},\langle m ^{+h(i)},h \rangle \models \varphi \mbox{ for every } i\in \mathds{N}\\

\mathcal{M},\langle m,h \rangle \models  \phi\mathsf{U} \psi &
\mbox{iff}& \emph{there is } j\in \mathds{N} \mbox{ s.t. } \mathcal{M},\langle m ^{+h(j)},h \rangle \models \psi \mbox{ and} \\ && \mathcal{M},\langle m ^{+h(i)},h \rangle \models \varphi\mbox{ for every } 0\leq i< j\\

\mathcal{M}, \langle m,h \rangle \models [C]
\varphi &\mbox{iff}& \mbox{for all }  h'\in \mathbf{\mathbf{Choice}}^m_{C}(h),  \mathcal{M},\langle m, h'\rangle \models \varphi\\

\mathcal {M},\langle m,h \rangle \models \langle\langle C\rangle\rangle^{s}
\phi & \mbox{ iff } & \mbox{there is a collective strategy } \sigma_C \mbox{ starting at $m$} \\&& \mbox{s.t. } \mathcal {M},\langle m, h'\rangle \models \phi \mbox{ for every } h'\in \mathbf{Adm}_C^m\left(\sigma_C\right).
\end{array} \]
\normalsize
Satisfiability, validity on a frame, and general validity are defined as usual. 
\end{definition}

On the path to proving a correspondence result, let us define the labelled \emph{bdt}-frame associated to a CGS.

\sloppy
\begin{definition}[Labelled \emph{bdt}-frame associated to a CGS]
\label{lbdtasscgs}

Let $\mathcal{S}=\langle W, Ags, Act, \delta,\mathcal{V}\rangle$ be a CGS. A structure $\mathcal{M}^{\mathcal{S}}=\langle M^{\mathcal{S}}, \sqsubset, Ags, \mathbf{Choice}, Tps, Lbl, Exe, \mathcal{V}^\mathcal{S} \rangle$ is defined as follows:
\begin{itemize}
    \item For each $w\in W$, let $out(w):=\bigcup_{f_{Ags}\in Strat_{Ags}}out(w,f_{Ags})^{fin}$. 
    
    We set $M^{\mathcal{S}}=\bigcup_{w\in W}out(w)$. Observe that this is a disjoint union, by definition. 
    
    This means that each moment in $M^{\mathcal{S}}$ is a finite sequence $\lambda\in W^+$ such that $\lambda(1)= w$ for some $w\in W$ and $\lambda(i+1)=\lambda(i)^{+f_{Ags}\left(\lambda[1,i]\right)}$ for every $1\leq i < len(\lambda)$---for some particular collective strategy $f_{Ags}\in Strat_{Ags}$, where recall that  $f_{Ags}\left(\lambda[1,i]\right)=\left\langle f_\alpha\left(\lambda[1,i]\right)\right\rangle_{\alpha\in Ags}$ is an action profile in $Act_{Ags}^{\lambda(i)}$.  
    
    \item $\sqsubset$ on $M^{\mathcal{S}}$ is defined by the following rule: 
    
    $\lambda\sqsubset \lambda'$ iff $len(\lambda)<len(\lambda')$ and $\lambda'[1,len(\lambda)]=\lambda$ (i.e., for all $1\leq i\leq len(\lambda)$, $\lambda'(i)=\lambda(i)$).
    
   The set $H\subseteq 2^{M}$ of \emph{histories} is defined as usual: each history is a maximal set of linearly ordered moments. Observe that, for each $h\in H$, $h$ coincides with the set $out\left(w,f_{Ags}\right)^{fin}$ for some $f_{Ags}$, and that it is the case that $h= out\left(w,f_{Ags}'\right)^{fin}$ for every $f_{Ags}'\in Strat_{Ags}$ such that $f_{Ags}'(\lambda)=f_{Ags}(\lambda)$ for every $\lambda\in h$.\footnote{\label{sol}Let us show that for each $h\in H$, $h=out\left(w, f_{Ags}\right)^{fin}$ for some $w\in W$ and $f_{Ags}\in Strat_{Ags}$. Let $h\in H$. First of all, observe that, by definition of $\sqsubset$, it is the case that for each $i\in \mathds{N}-\{0\}$, there is exactly one moment $\lambda_i\in h$ such that $len(\lambda_i)=i$. Furthermore, for each $i\in \mathds{N}-\{0\}$, by definition of $M^{\mathcal{S}}$, it is the case that $\lambda_i\in out\left(\lambda_1(len(\lambda_1)), h_{Ags}^{\lambda_i}\right)^{fin}$ for some $h_{Ags}^{\lambda_i}\in Strat_{Ags}$. Let us define a collective strategy $f_{Ags}^h$ over $W^+$ by the following rules: for each $i\in \mathds{N}-\{0\}$, let $f_{Ags}^h(\lambda_i)=h_{Ags}^{\lambda_{i+1}}(\lambda_i)$; for every other sequence $\lambda\in W^+$, let $f_{Ags}^h(\lambda)$ be any collective strategy in $Act_{Ags}^{\lambda(len(\lambda))}$. Using induction on the length of sequences, one can easily show that $h=out\left(\lambda_1(len(\lambda_1)), f_{Ags}^h\right)^{fin}$.}
   Therefore, each $h\in H$ is associated to a single sequence of action profiles given by $\{f_{Ags}(\lambda)\}_{\lambda\in h}$. 
   
   Observe that every state $w$ of $\mathcal{S}$ is therefore associated to an unraveled tree, given by the tuple $\langle out(w), \sqsubset \rangle$. If we take $\lambda_w$ to be the \emph{sequence} ``$w$'' (for which $len(\lambda_w)= 1$ and $\lambda_w(1)=w$), then it is clear that $H_{\lambda_w}=\{h\in H; min(h)=\lambda_w\}$. Furthermore, we have that $H=\bigcup_{w\in W}H_{\lambda_w}$ (where this is a disjoint union). Therefore, every history $h$ in $H$ has a (unique) root moment, which is denoted by $min(h)$. 
   
   For $h\in H$, I will denote by $seq(h)$ the infinite \emph{sequence} given by the elements of $h$. Therefore, for each $h\in H$, $seq(h)=out\left(w, f_{Ags}^h\right)$.  
   
    For each $\lambda\in M^{\mathcal{S}}$ (with $\lambda(1)=w$) and each history $h\in H_\lambda$ (which implies that  $min(h)=\lambda_w$), there exists a unique $\lambda'$, known as $\lambda$'s successor along $h$, such that $\lambda \sqsubset \lambda'$ and $\lambda'\sqsubseteq \lambda''$ for every $\lambda''$ such that $\lambda \sqsubset \lambda''$. The proof of this property, which I have referred to as \emph{time-discreteness} is included below, in the first item of the proof of Proposition \ref{indeed}. For each $\lambda$ and each history $h\in H_\lambda$, I will denote $\lambda$'s successor along $h$ by $\lambda^{+h}$.   
   
   For each history $h\in H$ (with $h=out(w,f_{Ags}^h)^{fin}$ and $min(h)=\lambda_w$), the single sequence of action profiles associated to $h$ can be therefore ordered thus:
   
   $f_{Ags}^h(\lambda_w), f_{Ags}^h\left((\lambda_w)^{+h}\right),  f_{Ags}^h\left(\left((\lambda_w)^{+h}\right)^{+h}\right)\dots$. For each history $h\in H$, its associated sequence of action profiles will also be denoted in the form  $\mathbf{s}_1^h, \mathbf{s}_2^h, \mathbf{s}_3^h,\dots$ henceforward. Observe that it is the case that $\mathbf{s}_1^h\in Act_{Ags}^{w}$, $\mathbf{s}_2^h\in  Act_{Ags}^{w^{+\mathbf{s}_1^h}}$, $\mathbf{s}_3^h\in  Act_{Ags}^{(w^{+\mathbf{s}_1^h})^{+\mathbf{s}_{2}^h}}$, etc, and that for each $\lambda\in h$, $f_{Ags}^h(\lambda)=\mathbf{s}^h_{len(\lambda)}$, so that $\lambda^{+h}\left(len\left(\lambda^{+h}\right)\right)=\lambda\left(len(\lambda)\right)^{+\mathbf{s}_{len(\lambda)}^h}$.\footnote{\label{bonham}Observe that, for $\lambda\in M^{\mathcal{S}}$, the partition $\mathbf{Choice}_{Ags}^\lambda$ is in a one-to-one correspondence with $Act_{Ags}^{\lambda\left(len (\lambda)\right)}$ (for $h, h'\in H_{\lambda}$, $\mathbf{s}^h_{len(\lambda)}= \mathbf{s}^{h'}_{len(\lambda)}$ iff $\mathbf{Choice}_{Ags}(h)=\mathbf{Choice}_{Ags}(h')$). Furthermore, it is the case that $\mathbf{s}^h_{len(\lambda)}= \mathbf{s}^{h'}_{len(\lambda)}$ iff $\lambda^{+h}=\lambda^{+h'}$. This last quality is due to the fact that, in $\mathcal{S}$, each action profile corresponds to a unique successor state, as implied by the definition of the transition function $\delta$.} 

    \item Recall that $\lambda\left(len(\lambda)\right)$ is the last state in the sequence $\lambda$, and that $\delta[w, \mathbf{s}_\alpha]$ is defined as the set of all the possible successor states that may result from executing $\alpha$'s action $\mathbf{s}_\alpha$.  $\mathbf{Choice}$ is defined by the following rule. For $\alpha\in Ags$ and $\lambda\in M^{\mathcal{S}}$, we set
    \[\mathbf{Choice}_\alpha^\lambda(h):=\left\{h'\in H; \lambda^{+h'}\left(len\left(\lambda^{+h'}\right)\right)\in \delta\left[\lambda\left(len\left(\lambda\right)\right), \left(\mathbf{s}^h_{len\left(\lambda\right)}\right)_\alpha\right] \right\}.\]
    
    Observe that this definition implies that $h'\in \mathbf{Choice}_\alpha^\lambda(h)$ iff $\left(\mathbf{s}^h_{len\left(\lambda\right)}\right)_\alpha=\left(\mathbf{s}^{h'}_{len\left(\lambda\right)}\right)_\alpha$.

    \item $Tps$ is defined as the set $\bigcup_{\alpha\in Ags, w\in W}Act_\alpha^w$, such that, for $\alpha\in Ags$ and $\lambda\in M^{\mathcal{S}}$, $Tps_\alpha^\lambda=Act_\alpha^{\lambda\left(len(\lambda)\right)}$.
    
    \item $Lbl$ is defined as follows. For $\alpha\in Ags$,  $\lambda\in M^{\mathcal{S}}$, and $L\in \mathbf{Choice}_\alpha^\lambda$, $Lbl(L)=\left(\mathbf{s}^h_{len\left(\lambda\right)}\right)_\alpha$, where $h$ is any history within $L$. 
    
    Observe that this definition, with the one of $\mathbf{Choice}$ above, implies that $h'\in \mathbf{Choice}_\alpha^\lambda(h)$ iff $Lbl\left(\mathbf{Choice}_\alpha^\lambda(h')\right)=Lbl\left(\mathbf{Choice}_\alpha^\lambda(h)\right)$.
    
    \item $Exe$ is defined as follows. For $\alpha\in Ags$,  $\lambda\in M^{\mathcal{S}}$, and $\tau\in Tps$ such that $\tau\in Tps_\alpha^\lambda$ (which implies that $\tau=s_{\alpha,\lambda\left(len\left(\lambda\right)\right)}$ for some action label $s_{\alpha,\lambda\left(len(\lambda)\right)}\in Act_\alpha^{\lambda\left(len(\lambda)\right)}$),   $Exe_\alpha^\lambda(\tau)=\left\{h\in H_\lambda; \left(\mathbf{s}^h_{len\left(\lambda\right)}\right)_\alpha=  s_{\alpha,\lambda\left(len(\lambda)\right)}\right\}$. Equivalently, in terms of the labelling function defined above, $Exe_\alpha^\lambda(\tau)=\left\{h\in H_\lambda; Lbl\left(\mathbf{Choice}_\alpha^\lambda(h)\right)=  s_{\alpha,\lambda\left(len(\lambda)\right)}\right\}$.
    
    \item $\mathcal{V}^\mathcal{S}:P\to 2^{I(M^{\mathcal{S}}\times H)}$ is defined by  the following rule: $\mathcal{V}^\mathcal{S}(p)=\left\{\langle \lambda, h\rangle;\lambda\left(len(\lambda)\right)\in \mathcal{V}(p)\right\} $.
\end{itemize}
\end{definition}

\begin{proposition}
\label{indeed}

Let $\mathcal{S}=\langle W, Ags, Act, \delta,\mathcal{V}\rangle$ be a concurrent game structure. The structure $\mathcal{M}^{\mathcal{S}}$---as defined in Definition \ref{lbdtasscgs}---is indeed a deterministic labelled \emph{bdt}-frame.
\end{proposition}

\begin{proof} We need to show that $\mathcal{M}^{\mathcal{S}}$ meets the conditions stated in Definition \ref{labelbdt}. 
\begin{itemize}
    \item Let us show that $\langle M^{\mathcal{S}}, \sqsubset, Ags, \mathbf{Choice},  \mathcal{V}^\mathcal{S} \rangle$ is a \emph{bdt}-model: \begin{itemize}
        \item The definition of $len$ and of $\sqsubset$ implies straightforwardly that $\sqsubset$ is irreflexive and transitive. The fact that $<$ is trichotomous over $\mathds{N}-\{0\}$ implies that $\sqsubset$ satisfies no backward branching. 
        
        As for the condition of \emph{time-discreteness}, let $\lambda\in M^{\mathcal{S}}$ and $h\in H_\lambda$ (with $\lambda(1)=w$ and $h\in out(w, f_{Ags})$). Let $\lambda'$ be the sequence in $W^+$ such that $len(\lambda')=len(\lambda)+1$,  $\lambda'[1,len(\lambda)]=\lambda$, and $\lambda'(len(\lambda'))=\left(\lambda\left(len(\lambda)\right)\right)^{+f_{Ags}(\lambda)}$. 
        
        It is clear that $\lambda\sqsubset \lambda'$. Now assume that $\lambda \sqsubset \lambda''$ for some $\lambda''\in h$. This means that $len(\lambda'')>len(\lambda)$. If $len(\lambda'')=len(\lambda)+1$, then, by definition of $\sqsubset$, we have that $\lambda''=\lambda'$. If $len(\lambda'')>len(\lambda)+1$, then it is clear that $len(\lambda'')>len(\lambda')$, so that the same definition of $\sqsubset$ and the assumption that $\lambda''\in h$ implies that $\lambda'\sqsubset \lambda''$. Therefore $\lambda'$ is the successor of $\lambda$ along $h$, which is denoted by $\lambda^{+h}$.

        \item As for $\mathbf{Choice}$, let $\alpha\in Ags$ and $\lambda\in M^{\mathcal{S}}$. Recall from Footnote \ref{bonham} that every $h\in H_{\lambda}$ corresponds to a unique action profile $\mathbf{s}_{len(\lambda)}^h\in Act_{Ags}^{\lambda\left(len(\lambda)\right)}$. Therefore, $\bigcup \mathbf{Choice}_\alpha^\lambda=H_\lambda$. Now, if  $\mathbf{Choice}_\alpha^\lambda(h)\neq \mathbf{Choice}_\alpha^\lambda(h')$, then, by the observation immediately after the definition of $\mathbf{Choice}_\alpha^\lambda(h)$ in Definition \ref{lbdtasscgs}, it is the case that $\left(\mathbf{s}^h_{len\left(\lambda\right)}\right)_\alpha\neq \left(\mathbf{s}^{h'}_{len\left(\lambda\right)}\right)_\alpha$. The assumption that there is a history $h''$ in the intersection of these two cells would imply---by the same observation---that $\left(\mathbf{s}^h_{len\left(\lambda\right)}\right)_\alpha= \left(\mathbf{s}^{h''}_{len\left(\lambda\right)}\right)_\alpha=\ \left(\mathbf{s}^{h'}_{len\left(\lambda\right)}\right)_\alpha$, which is a contradiction. Therefore $\mathbf{Choice}_\alpha^\lambda$ is indeed a partition of $H_\lambda$. 
        
        Let us show that $\mathbf{Choice}$ satisfies condition $(\mathtt{NC})$. Let $h, h'\in H_\lambda$ such that there exists $\lambda'\sqsupset \lambda$ such that $\lambda'\in h\cap h'$. Observe that $\lambda^{+h}\sqsubseteq \lambda'$ and $\lambda^{+h'}\sqsubseteq \lambda'$, which implies that $\lambda^{+h}\left(len\left(\lambda^{+h}\right)\right)=\lambda'\left(len\left(\lambda^{+h}\right)\right)=\lambda^{+h'}\left(len\left(\lambda^{+h'}\right)\right)$. Since additionally $\lambda^{+h}[1, len\left(\lambda^{+h}\right)-1]=\lambda=\lambda^{+h'}[1, len\left(\lambda^{+h'}\right)-1]$, it is the case that $\lambda^{+h}=\lambda^{+h'}$. Therefore, $\mathbf{s}_{len(\lambda)}^h=\mathbf{s}_{len(\lambda)}^{h'}$, which straightforwardly implies that $h,h'$ lie within the same cell in $\mathbf{Choice}_\alpha^\lambda$.
        
        Let us show that $\mathbf{Choice}$ satisfies condition $(\mathtt{IA})$. Let $s\in \mathbf{Select}^\lambda$ be a selection function such that $h_\alpha\in s(\alpha)$ for every $\alpha\in Ags$. The action profile given by $\left\langle \left(\mathbf{s}^{h_\alpha}_{len\left(\lambda\right)}\right)_\alpha \right\rangle_{\alpha\in Ags}$ clearly lies in $Act_{Ags}^{\lambda\left(len(\lambda)\right)}$. By the observation in Footnote \ref{bonham}, there exists a history $h_*\in H_\lambda$ corresponding to this action profile---i.e., such that $\mathbf{s}^{h_*}_{len(\lambda)}=\left\langle \left(\mathbf{s}^{h_\alpha}_{len\left(\lambda\right)}\right)_\alpha \right\rangle_{\alpha\in Ags}$. By definition, $h_*\in \bigcap_{\alpha\in Ags}s(\alpha)$, then. Furthermore, Footnote \ref{bonham} implies that for every $h'\in \bigcap_{\alpha\in Ags}s(\alpha)$, $m^{+h'}=m^{+h_*}$. It is because of this condition that $\mathcal{M}$ is called a \emph{deterministic} \emph{bdt}-model. 
        
        \item $\mathcal{V}^\mathcal{S}$ is a well-defined valuation function. 
    \end{itemize}
    \item $Tps$ is well defined. 
    \item $Lbl$ is well defined. 
    \item $Exe$ is well defined, so that the observation immediately after the definition of $\mathbf{Choice}_\alpha^\lambda(h)$ in Definition \ref{lbdtasscgs} implies that for all $\tau\in Tps_\alpha^\lambda$ ($\lambda\in M^\mathcal{S}$, $\alpha \in Ags$), $Exe_\alpha^\lambda(\tau)\in \mathbf{Choice}_\alpha^\lambda$. 
    \begin{itemize}
        \item As for $(\mathtt{EL})$, observe that $h'\in Exe\left(Lbl(\mathbf{Choice}^\lambda_\alpha(h))\right)$ iff

        $\left(\mathbf{s}^{h'}_{len\left(\lambda\right)}\right)_\alpha=Lbl\left(\mathbf{Choice}_\alpha^\lambda(h)\right)=\left(\mathbf{s}^{h'}_{len\left(\lambda\right)}\right)_\alpha$ iff $h'\in \mathbf{Choice}^\lambda_\alpha(h)$ (where this last double implication occurs in virtue of the observation immediately after the definition of $\mathbf{Choice}_\alpha^\lambda(h)$ in Definition \ref{lbdtasscgs}). 
        
        \item As for $(\mathtt{LE})$, the definition of $Lbl$ and $Exe$ straightforwardly imply that $Lbl\left( Exe_\alpha^\lambda(\tau)\right)=\tau$.
    \end{itemize}
\end{itemize}
Therefore, $\mathcal{M}$ is indeed a deterministic labelled \emph{bdt}-model. 
\end{proof}

For the correspondence between labelled \emph{bdt}-models and concurrent game structures, we define a translation:

\begin{definition}[Translation]\label{def_translation2}
Let $p\in\mathcal{P}$ be an atomic proposition, let $\varphi,\psi\in\mathcal{L}_{\textsf{ATL}}$, and let $C\subseteq Ags$ be a coalition. A translation from $\mathcal{L}_{\textsf{ATL}}$ to $\mathcal{L}_{\textsf{SX}}$ is defined recursively as follows:
\[\begin{array}{lll}
Tr(p)&=& \square p\\

Tr(\neg\varphi)&=& \neg Tr(\varphi)\\

Tr(\varphi\wedge\psi)&=& Tr(\varphi)\wedge Tr(\psi)\\

Tr\left(\langle\langle C\rangle\rangle X\phi\right)&=&\langle\langle C\rangle\rangle^s X Tr(\varphi)\\

Tr\left(\langle\langle C\rangle\rangle \mathsf{G}\phi\right)&=&\langle\langle C\rangle\rangle^s \mathsf{G} Tr(\varphi)\\

Tr\left(\langle\langle C\rangle\rangle \phi\mathsf{U}\psi\right)&=&\langle\langle C\rangle\rangle^s Tr(\varphi)\mathsf{U} Tr(\psi).
\end{array}\]
\end{definition}

\fussy

\begin{lemma}
\label{dersicola0} Let $\lambda, \lambda'\in M^{\mathcal{S}}$ such that $\lambda(len(\lambda))=\lambda'(len(\lambda'))$. There exists a one-to-one correspondence $c:H_\lambda\to H_{\lambda'}$ such that, for every $h\in H_\lambda$ and $i\in \mathds{N}$, it is the case that $\mathbf{s}_{len(\lambda)+i}^h=\mathbf{s}_{len(\lambda')+i}^{c(h)}$.
\end{lemma}
\begin{proof}
Follows straightforwardly from Footnote \ref{bonham} and the paragraph it annotates in Definition \ref{lbdtasscgs}.
\end{proof}

\begin{remark}\label{dedogordo}
     
    Observe that the set $\{\mu'\in h; \mu'\sqsupseteq \mu\}$ can be rewritten in the form $\{\mu^{+h(i)}\}_{i\in \mathds{N}}$. For $\lambda, \lambda'\in M^{\mathcal{S}}$ such that $\lambda(len(\lambda))=\lambda'(len(\lambda'))$, then, the correspondence $c$ between $H_\lambda$ and $H_{\lambda'}$ ensures that, for each $i\in \mathds{N}$,  $\lambda^{+h(i)}\left(len\left(\lambda^{+h(i)}\right)\right)=\lambda'^{+c(h)(i)}\left(len\left(\lambda'^{+c(h)(i)}\right)\right)$. This is due to the fact that each action profile determines a unique successor state in $\mathcal{S}$.

\end{remark}

\begin{lemma}
\label{dersicola} Let $\phi$ be a formula of $\mathcal{L}_{\textsf{SX}}$, and let $\lambda, \lambda'\in M^{\mathcal{S}}$ such that $\lambda(len(\lambda))=\lambda'(len(\lambda'))$, where observe that $\mathcal{V}^\mathcal{S}$ specifies the valuation of atomic propositions in $\mathcal{M}^\mathcal{S}$. For each $h\in H_\lambda$, $\mathcal{M}^{\mathcal{S}},\langle \lambda, h\rangle\models \phi$ iff $\mathcal{M}^{\mathcal{S}},\langle \lambda', c(h)\rangle\models \phi$. 
\end{lemma}
\begin{proof}

Let $h\in H_\lambda$. We proceed by induction on the complexity of $\phi$. The base case is immediate from the definition of $\mathcal{V}^\mathcal{S}$ in Definition \ref{lbdtasscgs}. The cases with the Boolean connectives are standard. Let us deal with the cases for the remaining modal formulas.
\begin{itemize}
    \item (``$X\phi$'')
    
    The correspondence $c$ between $H_\lambda$ and $H_{\lambda'}$ ensures that $\lambda^{+h}\left(len \left(\lambda^{+h}\right)\right)=\lambda'^{+c(h)}\left(len \left(\lambda'^{+c(h)}\right)\right)$. It is the case that $\mathcal{M}^{\mathcal{S}},\langle \lambda, h\rangle\models X \phi$ iff $\mathcal{M}^{\mathcal{S}},\langle \lambda^{+h}, h\rangle\models \phi$, which, by induction hypothesis and the above observation regarding the correspondence $c$, occurs iff $\mathcal{M}^{\mathcal{S}},\langle \lambda'^{+c(h)}, c(h)\rangle\models \phi$, which in turn occurs iff $\mathcal{M}^{\mathcal{S}},\langle \lambda', c(h)\rangle\models X\phi$.
    
    \item (``$\mathsf{G}\phi$'')
    
    $\mathcal{M}^{\mathcal{S}},\langle \lambda, h\rangle\models \mathsf{G} \phi$ iff for all $i\in \mathds{N}$, $\mathcal{M}^{\mathcal{S}},\left\langle \lambda^{+h(i)}, h\right\rangle\models \phi$, which, by induction hypothesis and Remark \ref{dedogordo}, occurs iff for all $i\in \mathds{N}$, $\mathcal{M}^{\mathcal{S}},\left\langle \lambda'^{+c(h)(i)}, c(h)\right\rangle\models \phi$, which in turn occurs iff $\mathcal{M}^{\mathcal{S}},\langle \lambda', c(h)\rangle\models \mathsf{G} \phi$.
    
    \item (``$\phi\mathsf{U}\psi$'')
    
   $\mathcal{M}^{\mathcal{S}},\langle \lambda, h\rangle\models \phi\mathsf{U}\psi$ iff there exists $j\in \mathds{N}$ such that 
    $\mathcal{M}^{\mathcal{S}}, \left\langle \lambda^{+h(j)}, h\right\rangle\models \psi$ and $\mathcal{M}^{\mathcal{S}},\left\langle \lambda^{+h(i)}, h\right\rangle\models \phi$ for every $0\leq i<j$, which, by induction hypothesis, occurs iff there exists $j\in \mathds{N}$ such that $\mathcal{M}^{\mathcal{S}},\left\langle \lambda'^{+c(h)(j)}, c(h)\right\rangle\models \psi$ and  $\mathcal{M}^{\mathcal{S}},\left\langle \lambda'^{+c(h)(i)},c(h)\right\rangle\models \phi$ for every  $0\leq i<j$, which in turn occurs iff $\mathcal{M}^{\mathcal{S}},\langle \lambda', c(h)\rangle\models \phi\mathsf{U}\psi$.
    
    \item (``$\square\phi$'')
    
    $\mathcal{M}^{\mathcal{S}},\langle \lambda, h\rangle\models \square \phi$ iff for every $h'\in H_{\lambda}$, $\mathcal{M}^{\mathcal{S}},\langle \lambda, h'\rangle\models \phi$,  which, by induction hypothesis, occurs iff for every $h'\in H_{\lambda}$ $\mathcal{M}^{\mathcal{S}},\langle \lambda', c(h')\rangle\models \phi$, which---in virtue of the fact that $c$ is a one-to-one correspondence---occurs iff $\mathcal{M}^{\mathcal{S}},\langle \lambda', c(h)\rangle\models \square\phi$.
    
    \item (``$\langle\langle C\rangle\rangle^s\phi$'')
    
    Observe that the correspondence $c$ induces a correspondence between $\sqsubseteq[\lambda]$ and $\sqsubseteq[\lambda']$, such that, for each $i\in\mathds{N} $, $\lambda^{+h(i)}\left(len\left(\lambda^{+h(i)}\right)\right)=\lambda'^{+c(h)(i)}\left(len\left(\lambda'^{+c(h)(i)}\right)\right)$. Therefore, for each collective strategy $\sigma_C$ starting at $\lambda$, there is a corresponding collective strategy $\sigma_C'$ starting at $\lambda'$ such that, for each $h\in H_\lambda$ and $i\in \mathds{N}$, $\sigma_C\left( \lambda^{+h(i)}\right)=\sigma_C'\left( \lambda^{+c(h)(i)}\right)$, and viceversa. Furthermore, $h'\in \mathbf{Adm}_C^\lambda\left(\sigma_C\right)$ iff $c(h')\in \mathbf{Adm}_C^{\lambda'}\left(\sigma_C'\right)$.  
    
    
     Therefore, it is the case that $\mathcal{M}^{\mathcal{S}},\langle \lambda, h\rangle\models \langle\langle C\rangle\rangle^s\phi$ iff there exists $\sigma_C$ such that, for every $h'\in \mathbf{Adm}_C^\lambda\left(\sigma_C\right)$, $\mathcal{M}^{\mathcal{S}},\langle \lambda, h'\rangle\models \phi$,  which, by induction hypothesis and the observation above, occurs iff there exists $\sigma_C'$ such that, for every $h''\in \mathbf{Adm}_C^{\lambda'}\left(\sigma_C'\right)$, $\mathcal{M}^{\mathcal{S}},\langle \lambda', h''\rangle\models \phi$, which in turn occurs iff $\mathcal{M}^{\mathcal{S}},\langle \lambda', c(h)\rangle\models \langle\langle C\rangle\rangle^s\phi$.
\end{itemize}

\end{proof}

\begin{lemma}
\label{dersicola1} Let $\phi$ be a formula of $\mathcal{L}_{\textsf{ATL}}$, then for a given $\lambda\in M^{\mathcal{S}}$ and $h\in H_\lambda$, $\mathcal{M}^{\mathcal{S}}, \langle \lambda, h \rangle\models Tr(\phi)$ iff $\mathcal{M}^{\mathcal{S}}, \langle \lambda, h' \rangle\models Tr(\phi)$ for every $h'\in H_\lambda$. 
\end{lemma}
\begin{proof}

Straightforward. It can be shown by induction on the complexity of $\phi$, where the cases for the modal formulas follow from the fact that, for $\lambda\in M^{\mathcal{S}}$, the existence of strategies starting at $\lambda$ in $\mathcal{M}^{\mathcal{S}}$ does not depend on any history passing through $\lambda$. 

\end{proof}

\begin{restatable}[Correspondence]{proposition}{corr}
\label{gottabe}
 Let $\mathcal{S}=\langle W, Ags, Act, \delta,\mathcal{V}\rangle$ be a concurrent game structure such that $\mathcal{M}^{\mathcal{S}}$ is the labelled \emph{bdt}-frame associated to $\mathcal{S}$. For each formula $\phi$ of $\mathcal{L}_{\textsf{ATL}}$ and state $w\in W$, it is the case that $\mathcal{S},w\models \phi \mbox{ iff } \mathcal{M}^{\mathcal{S}},\langle\lambda_w,h\rangle\models Tr(\phi) \mbox{ for every } h\in H_{\lambda_w}$ (where $\lambda_w$ is a moment in $\mathcal{M}^{\mathcal{S}}$ that corresponds to state $w$.)
\end{restatable}

\begin{proof}
We proceed by induction on the complexity of $\phi$. The base case is immediate from Definition \ref{lbdtasscgs}. The cases with the Boolean connectives are standard. Let us deal with the cases for the remaining modal formulas:

\begin{itemize}
    \item (``$\langle\langle C\rangle\rangle X\phi$'')   
    
    ($\Rightarrow$) Assume that $\mathcal{S}, w\models \langle\langle C\rangle\rangle X\phi$. This means that there exists a collective strategy $f_C$ over $\mathcal{S}$ such that, for every $\mu\in out(w, f_C)$, $\mathcal{S}, \mu(2)\models \phi$. We want to show that there exists a collective strategy $\sigma_C$ starting at $\lambda_w$ such that, for each $h'\in \mathbf{Adm}_C^{\lambda_w}\left(\sigma_C\right)$, $\mathcal{M}^{\mathcal{S}},\langle\lambda_w,h'\rangle\models XTr(\phi)$.
    
    Let us define the collective strategy $\sigma_C$ starting at $\lambda_w$ that we need to ensure that this direction holds.  For each $\mu\in M^{\mathcal{S}}$ such that $\mu\sqsupseteq \lambda_w$, let $\sigma_C\left(\mu\right)=\bigcap_{\alpha\in C}Exe\left(\left(f_C(\mu)\right)_\alpha\right)$, where recall that $f_C$ is a tuple of the form $\langle f_\alpha \rangle_{\alpha\in C}$ such that $f_\alpha\in Strat_\alpha$ for every $\alpha\in C$, and that for each $\mu\in W^+$, $f_{C}(\mu)$ denotes the joint action---in $\mathcal{S}$---of $Act_{C}^{\mu(len(\mu))}$ given by $\langle f_\alpha(\mu) \rangle_{\alpha\in C}$.  
    
    
     Let $h'\in \mathbf{Adm}_C^{\lambda_w}\left(\sigma_C\right)$, which implies that $Lbl\left(\mathbf{Choice}_\alpha^{\lambda_w^{+h'(i)}}(h')\right)=\left(f_C\left(\lambda_w^{+h'(i)}\right)\right)_\alpha$ for every $\alpha\in C$ and $i\in \mathds{N}$. In turn, this implies that $\left(f_{Ags}^{h'}\left(\lambda_w^{+h'(i)}\right)\right)_\alpha=\left(f_C\left(\lambda_w^{+h'(i)}\right)\right)_\alpha$ for every $\alpha\in C$ and $i\in\mathds{N}$. Therefore, $seq(h')\in out(w, f_C)$. By assumption $\mathcal{S}, seq(h')(2)\models \phi$.

    
    By induction hypothesis, $\mathcal{M}^{\mathcal{S}}, \left\langle \lambda_{seq(h')(2)}, h'' \right\rangle \models Tr(\phi)$ for any history $h''\in H_{\lambda_{seq(h')(2)}}$, where recall that $\lambda_{seq(h')(2)}$ is the \emph{sequence} whose first and only element is $seq(h')(2)=w^{+\mathbf{s}_1^{h'}}=w^{+f_{Ags}^{h'}(\lambda_w)}=\lambda_w^{+h'}\left(len\left(\lambda_w^{+h'}\right)\right)$.  Lemma \ref{dersicola} then implies that $\mathcal{M}^{\mathcal{S}}, \left\langle (\lambda_w)^{+h'}, h' \right\rangle \models Tr(\phi)$. 
    
    Therefore, $\mathcal{M}^{\mathcal{S}}, \langle \lambda_w, h' \rangle \models XTr(\phi)$. Since $h'$ was an arbitrary history in $\mathbf{Adm}_C^{\lambda_w}\left(\sigma_C\right)$, it is the case that  $\mathcal{M}^{\mathcal{S}},\langle\lambda_w,h\rangle\models \langle\langle C\rangle\rangle^sXTr(\phi)$ for any $h\in H_{\lambda_w}$ (where this last universal quantification over the histories in $H_{\lambda_w}$ follows from Lemma \ref{dersicola1}). 
    
    ($\Leftarrow$) Assume that $\mathcal{M}^{\mathcal{S}},\langle\lambda_w,h\rangle\models \langle\langle C\rangle\rangle^sXTr(\phi)$ for every $h\in H_{\lambda_w}$. This means that there exists a collective strategy $\sigma_C$ starting at $\lambda_w$ such that, for every $h'\in \mathbf{Adm}_C^{\lambda_w}\left(\sigma_C\right)$ $\mathcal{M}^{\mathcal{S}},\langle\lambda_w,h'\rangle\models XTr(\phi)$. We want to show that there exists a collective strategy $f_C$ over $\mathcal{S}$ such that, for every $\mu\in out(w, f_C)$, $\mathcal{S}, \mu(2)\models \phi$. 
    
    Let us build the collective strategy $f_C$ over $\mathcal{S}$ that we need to ensure that this direction holds. Let $f_C:W^+\to \bigcup_{w\in W}Act_\alpha^w$ be defined as follows. For each $h\in H_{\lambda_w}$, $i\in \mathds{N}$,  and $\alpha\in C$, let $\left(f_C\left(\lambda_w^{+h(i)}\right)\right)_\alpha =Lbl\left(\left(\sigma_C\right)_\alpha\left(\lambda_w^{+h(i)}\right)\right)$; for every other $\mu\in W^+$, let  $\left(f_C\left(\mu\right)\right)_\alpha$ be any action in $Act_\alpha^{\mu(len(\mu))}$. 
    
    
    Let $\mu\in out(w,f_C)$, which implies that $\mu^{fin}\in H_{\lambda_w}$. Let us show that $\mu^{fin}\in \mathbf{Adm}_C^{\lambda_w}\left(\sigma_C\right)$. In order to do this, we need to prove that $\mu^{fin}\in \sigma_C\left(\lambda_w^{+\mu^{fin}(i)}\right)$ for every $i\in \mathds{N}$. Let $i\in \mathds{N}$. Recall that $\sigma_C\left(\lambda_w^{+\mu^{fin}(i)}\right)$ is a choice-cell in $\mathbf{Choice}_C^{\lambda_w^{+\mu^{fin}(i)}}$, given by the intersection $\bigcap_{\alpha\in C}\left(\sigma_C\right)_\alpha\left(\lambda_w^{+\mu^{fin}(i)}\right)$. Let $\alpha\in C$. By construction, the collective strategy $f_{Ags}^{\mu^{fin}}$ associated to history $\mu^{fin}$---where $\mu^{fin}=out\left(w, f_{Ags}^{\mu^{fin}}\right)^{fin}$---is such that $(\star)$ $\left(f_{Ags}^{\mu^{fin}}\left(\lambda_w^{+\mu^{fin}(i)}\right)\right)_\alpha =\left(f_C\left(\lambda_w^{+\mu^{fin}(i)}\right)\right)_\alpha$. 
    
    By the definitions of $\mathbf{Choice}$ and $Lbl$ in Definition \ref{lbdtasscgs}, it is the case that
    
    $Lbl\left(\mathbf{Choice}_\alpha^{\lambda_w^{+\mu^{fin}(i)}}\left(\mu^{fin}\right)\right)=\left(f_{Ags}^{\mu^{fin}}\left(\lambda_w^{+\mu^{fin}(i)}\right)\right)_\alpha$, so that the equality marked by $(\star)$ and the definition of $f_C$ imply that
    
    $Lbl\left(\mathbf{Choice}_\alpha^{\lambda_w^{+\mu^{fin}(i)}}\left(\mu^{fin}\right)\right)=Lbl\left(\left(\sigma_C\right)_\alpha\left(\lambda_w^{+\mu^{fin}(i)}\right)\right)$, which implies that $\mu^{fin}\in \left(\sigma_C\right)_\alpha\left(\lambda_w^{+\mu^{fin}(i)}\right)$. Therefore, $\mu^{fin}\in \sigma_C\left(\lambda_w^{+\mu^{fin}(i)}\right)$ for every $i\in \mathds{N}$, so that $\mu^{fin}\in \mathbf{Adm}_C^{\lambda_w}\left(\sigma_C\right)$. 
    
    By assumption, then, it is the case that $\mathcal{M}^{\mathcal{S}},\langle\lambda_w,\mu^{fin}\rangle\models XTr(\phi)$. Therefore, $\mathcal{M}^{\mathcal{S}},\left\langle\lambda_w^{+\mu^{fin}},\mu^{fin}\right\rangle\models Tr(\phi)$. It is clear that, for each $i\in \mathds{N}$, $\lambda_w^{+\mu^{fin}}\left(len\left(\lambda_w^{+\mu^{fin}}\right)\right)=\mu(2)$. Lemma \ref{dersicola} then implies that $\mathcal{M}^{\mathcal{S}},\left\langle\lambda_{\mu(2)},c\left(\mu^{fin}\right)\right\rangle\models Tr(\phi)$. In turn, Lemma \ref{dersicola1} implies that, for each $h\in H_{\lambda_{\mu(2)}}$, $\mathcal{M}^{\mathcal{S}},\left\langle\lambda_{\mu(2)},h\right\rangle\models Tr(\phi)$. By induction hypothesis, we get that $\mathcal{S},\mu(2)\models \phi$, so that $\mathcal{S}, w\models \langle\langle C\rangle\rangle X\phi$.
    
    
    \item (``$\langle\langle C\rangle\rangle \mathsf{G}\phi$'')
    
    ($\Rightarrow$)  Assume that $\mathcal{S}, w\models \langle\langle C\rangle\rangle X\phi$. This means that there exists a collective strategy $f_C$ over $\mathcal{S}$ such that, for every $\mu\in out(w, f_C)$, $\mathcal{S}, \mu(i)\models \phi$ for every $i\in \mathds{N}-\{0\}$. We want to show that there exists a collective strategy $\sigma_C$ starting at $\lambda_w$ such that, for each $h'\in \mathbf{Adm}_C^{\lambda_w}\left(\sigma_C\right)$, $\mathcal{M}^{\mathcal{S}},\langle\lambda_w,h'\rangle\models \mathsf{G}Tr(\phi)$. 
    
    Let us define the collective strategy $\sigma_C$ starting at $\lambda_w$ that we need to ensure that this direction holds. We do it just as in the right-to-left implication in the above item. For each $\mu\in M^{\mathcal{S}}$ such that $\mu\sqsupseteq \lambda_w$, let $\sigma_C\left(\mu\right)=\bigcap_{\alpha\in C}Exe\left(\left(f_C(\mu)\right)_\alpha\right)$. 
    
    Let $h'\in \mathbf{Adm}_C^{\lambda_w}\left(\sigma_C\right)$. Just as in the right-to-left implication of the above item, this implies that $seq(h')\in out(w, f_C)$. By assumption $\mathcal{S}, seq(h')(i)\models \phi$ for every $i\in \mathds{N}-\{0\}$. By induction hypothesis, it is the case that $(\star)$ $\mathcal{M}^{\mathcal{S}}, \left\langle \lambda_{seq(h')(i)}, h'' \right\rangle \models Tr(\phi)$ for every $i\in \mathds{N}-\{0\}$ and every history $h''\in H_{\lambda_{seq(h')(i)}}$. It is clear that, for each $i\in \mathds{N}$, $\lambda_w^{+h'(i)}\left(len\left(\lambda_w^{+h'(i)}\right)\right)=seq(h')(i+1)$. Therefore, Lemma \ref{dersicola} and $(\star)$ ensure that for every $i\in \mathds{N}$, $\mathcal{M}^{\mathcal{S}}, \left\langle \lambda_w^{+h'(i)}, h' \right\rangle \models Tr(\phi)$. 
    
    Therefore, $\mathcal{M}^{\mathcal{S}}, \left\langle \lambda_w^, h' \right\rangle \models \mathsf{G}Tr(\phi)$. Since $h'$ was an arbitrary history in $\mathbf{Adm}_C^{\lambda_w}\left(\sigma_C\right)$, it is the case that  $\mathcal{M}^{\mathcal{S}},\langle\lambda_w,h\rangle\models \langle\langle C\rangle\rangle^s\mathsf{G}Tr(\phi)$ for any $h\in H_{\lambda_w}$ (where this last universal quantification over the histories in $H_{\lambda_w}$ follows from Lemma \ref{dersicola1}).

    ($\Leftarrow$) Assume that $\mathcal{M}^{\mathcal{S}},\langle\lambda_w,h\rangle\models \langle\langle C\rangle\rangle^s\mathsf{G}Tr(\phi)$ for every $h\in H_{\lambda_w}$. This means that there exists a collective strategy $\sigma_C$ starting at $\lambda_w$ such that, for every $h'\in \mathbf{Adm}_C^{\lambda_w}\left(\sigma_C\right)$ $\mathcal{M}^{\mathcal{S}},\langle\lambda_w,h'\rangle\models \mathsf{G}Tr(\phi)$. We want to show that there exists a collective strategy $f_C$ over $\mathcal{S}$ such that, for every $\mu\in out(w, f_C)$, $\mathcal{S}, \mu(i)\models \phi$ for every $i\in \mathds{N}-\{0\}$.
    
     Let us build the collective strategy $f_C$ over $\mathcal{S}$ that we need to ensure that this direction holds. We do it just as in the left-to-right implication in the above item. Let $f_C:W^+\to \bigcup_{w\in W}Act_\alpha^w$ be defined as follows. For each $h\in H_{\lambda_w}$, $i\in \mathds{N}$,  and $\alpha\in C$, let $\left(f_C\left(\lambda_w^{+h(i)}\right)\right)_\alpha =Lbl\left(\left(\sigma_C\right)_\alpha\left(\lambda_w^{+h(i)}\right)\right)$; for every other $\mu\in W^+$, let  $\left(f_C\left(\mu\right)\right)_\alpha$ be any action in $Act_\alpha^{\mu(len(\mu))}$. 
     
     Let $\mu\in out(w,f_C)$, which implies that $\mu^{fin}\in H_{\lambda_w}$. Just as shown in the left-to-right implication in the above item, this implies that $\mu^{fin}\in \mathbf{Adm}_C^{\lambda_w}\left(\sigma_C\right)$.
     
     By assumption, it is the case that $\mathcal{M}^{\mathcal{S}},\langle\lambda_w,\mu^{fin}\rangle\models \mathsf{G}Tr(\phi)$. Therefore, $\mathcal{M}^{\mathcal{S}},\left\langle\lambda_w^{+\mu^{fin}(i)},\mu^{fin}\right\rangle\models Tr(\phi)$ for every $i\in\mathds{N}$. It is clear that, for each $i\in \mathds{N}$, $\lambda_w^{+\mu^{fin}(i)}\left(len\left(\lambda_w^{+\mu^{fin}(i)}\right)\right)=\mu(i+1)$. Lemma \ref{dersicola} then implies that $\mathcal{M}^{\mathcal{S}},\left\langle\lambda_{\mu(i)},c\left(\mu^{fin}\right)\right\rangle\models Tr(\phi)$ for every $i\in \mathds{N}-\{0\}$. Therefore, for each $i\in \mathds{N}-\{0\}$, Lemma \ref{dersicola1} implies that, for each $h\in H_{\lambda_{\mu(i)}}$, $\mathcal{M}^{\mathcal{S}},\left\langle\lambda_{\mu(i)},h\right\rangle\models Tr(\phi)$. By induction hypothesis, we get that $\mathcal{S},\mu(i)\models \phi$ for every $i\in \mathds{N}-\{0\}$, so that $\mathcal{S}, w\models \langle\langle C\rangle\rangle \mathsf{G}\phi$.
    
    \item (``$\langle\langle C\rangle\rangle \phi\mathsf{U}\psi$'')
    \sloppy
    ($\Rightarrow$)  Assume that $\mathcal{S}, w\models \langle\langle C\rangle\rangle \phi\mathsf{U}\psi$. This means that there exists a collective strategy $f_C$ over $\mathcal{S}$ such that, for every $\mu\in out(w, f_C)$, there exists $j\in \mathds{N}-\{0\}$ such that $\mathcal{S}, \mu(j)\models \psi$ and $\mathcal{S}, \mu(i)\models \phi$ for every $1\leq i<j$. We want to show that there exists a collective strategy $\sigma_C$ starting at $\lambda_w$ such that, for each $h'\in \mathbf{Adm}_C^{\lambda_w}\left(\sigma_C\right)$, $\mathcal{M}^{\mathcal{S}},\langle\lambda_w,h'\rangle\models Tr(\phi)\mathsf{U}Tr(\psi)$. 
    
    Let us define the collective strategy $\sigma_C$ starting at $\lambda_w$ that we need to ensure that this direction holds. We do it just as in the right-to-left implication in the above item. For each $\mu\in M^{\mathcal{S}}$ such that $\mu\sqsupseteq \lambda_w$, let $\sigma_C\left(\mu\right)=\bigcap_{\alpha\in C}Exe\left(\left(f_C(\mu)\right)_\alpha\right)$. 
    
    Let $h'\in \mathbf{Adm}_C^{\lambda_w}\left(\sigma_C\right)$. Just as in the right-to-left implication of the above item, this implies that $seq(h')\in out(w, f_C)$. By assumption, there exists $j\in \mathds{N}-\{0\}$ such that $\mathcal{S}, seq(h')(j)\models \psi$ and $\mathcal{S}, seq(h')(i)\models \phi$ for every $1\leq i<j$. 
    
    By induction hypothesis, it is the case that $(\star)$ $\mathcal{M}^{\mathcal{S}}, \left\langle \lambda_{seq(h')(j)}, h'' \right\rangle \models Tr(\psi)$ for every history $h''\in H_{\lambda_{seq(h')(j)}}$ and that $\mathcal{M}^{\mathcal{S}}, \left\langle \lambda_{seq(h')(i)}, h''' \right\rangle \models Tr(\phi)$ for every $1\leq i<j$ and  every history $h'''\in H_{\lambda_{seq(h')(i)}}$. It is clear that, for each $i\in \mathds{N}$, $\lambda_w^{+h'(i)}\left(len\left(\lambda_w^{+h'(i)}\right)\right)=seq(h')(i+1)$. Therefore, Lemma \ref{dersicola} and $(\star)$ ensure that $\mathcal{M}^{\mathcal{S}}, \left\langle \lambda_w^{+h'(j-1)}, h' \right\rangle \models Tr(\psi)$ and that $\mathcal{M}^{\mathcal{S}}, \left\langle \lambda_w^{+h'(i-1)}, h' \right\rangle \models Tr(\phi)$ for every $1\leq i<j$.
    
    Therefore, $\mathcal{M}^{\mathcal{S}}, \left\langle \lambda_w, h' \right\rangle \models Tr(\phi)\mathsf{U}Tr(\psi)$. Since $h'$ was an arbitrary history in $\mathbf{Adm}_C^{\lambda_w}\left(\sigma_C\right)$, it is the case that  $\mathcal{M}^{\mathcal{S}},\langle\lambda_w,h\rangle\models \langle\langle C\rangle\rangle^sTr(\phi)\mathsf{U}Tr(\psi)$ for any $h\in H_{\lambda_w}$ (where this last universal quantification over the histories in $H_{\lambda_w}$ follows from Lemma \ref{dersicola1}).

    ($\Leftarrow$) 
    
    Assume that $\mathcal{M}^{\mathcal{S}},\langle\lambda_w,h\rangle\models\langle\langle C\rangle\rangle^sTr(\phi)\mathsf{U}Tr(\psi)$ for every $h\in H_{\lambda_w}$. This means that there exists a collective strategy $\sigma_C$ starting at $\lambda_w$ such that, for every $h'\in \mathbf{Adm}_C^{\lambda_w}\left(\sigma_C\right)$ $\mathcal{M}^{\mathcal{S}},\langle\lambda_w,h'\rangle\models Tr(\phi)\mathsf{U}Tr(\psi)$. We want to show that there exists a collective strategy $f_C$ over $\mathcal{S}$ such that, for every $\mu\in out(w, f_C)$, there exists $j\in \mathds{N}_\{0\}$ such that $\mathcal{S}, \mu(j)\models \psi$ and $\mathcal{S}, \mu(i)\models \phi$ for every $1\leq i<j$.
    
    Let us build the collective strategy $f_C$ over $\mathcal{S}$ that we need to ensure that this direction holds. We do it just as in the left-to-right implication in the above item. Let $f_C:W^+\to \bigcup_{w\in W}Act_\alpha^w$ be defined as follows. For each $h\in H_{\lambda_w}$, $i\in \mathds{N}$,  and $\alpha\in C$, let $\left(f_C\left(\lambda_w^{+h(i)}\right)\right)_\alpha =Lbl\left(\left(\sigma_C\right)_\alpha\left(\lambda_w^{+h(i)}\right)\right)$; for every other $\mu\in W^+$, let  $\left(f_C\left(\mu\right)\right)_\alpha$ be any action in $Act_\alpha^{\mu(len(\mu))}$. 
     
     Let $\mu\in out(w,f_C)$, which implies that $\mu^{fin}\in H_{\lambda_w}$. Just as shown in the left-to-right implication in the above item, this implies that $\mu^{fin}\in \mathbf{Adm}_C^{\lambda_w}\left(\sigma_C\right)$.
     
     By assumption, it is the case that $\mathcal{M}^{\mathcal{S}},\langle\lambda_w,\mu^{fin}\rangle\models Tr(\phi)\mathsf{U}Tr(\psi)$. Therefore, there exists $j\in \mathds{N}$ such that $\mathcal{M}^{\mathcal{S}},\left\langle\lambda_w^{+\mu^{fin}(j)},\mu^{fin}\right\rangle\models Tr(\psi)$ and 
     
     $\mathcal{M}^{\mathcal{S}},\left\langle\lambda_w^{+\mu^{fin}(i)},\mu^{fin}\right\rangle\models Tr(\phi)$ for every $0\leq i<j$. It is clear that, for each $i\in \mathds{N}$, $\lambda_w^{+\mu^{fin}(i)}\left(len\left(\lambda_w^{+\mu^{fin}(i)}\right)\right)=\mu(i+1)$. Lemma \ref{dersicola} then implies that $\mathcal{M}^{\mathcal{S}},\left\langle\lambda_{\mu(j+1)},c\left(\mu^{fin}\right)\right\rangle\models Tr(\psi)$ and 
     $\mathcal{M}^{\mathcal{S}},\left\langle\lambda_{\mu(i+1)},c\left(\mu^{fin}\right)\right\rangle\models Tr(\phi)$ for every $0\leq i<j$.
     
     Lemma \ref{dersicola1} then implies that, for each $h\in H_{\lambda_{\mu(j+1)}}$, $\mathcal{M}^{\mathcal{S}},\left\langle\lambda_{\mu(j+1)},h\right\rangle\models Tr(\psi)$, and that, for each $0\leq i< j$ an each  each $h'\in H_{\lambda_{\mu(i+1)}}$, $\mathcal{M}^{\mathcal{S}},\left\langle\lambda_{\mu(i+1)},h'\right\rangle\models Tr(\phi)$. By induction hypothesis, we get that $\mathcal{S},\mu(j+1)\models \psi$ and $\mathcal{S},\mu(i+1)\models \psi$ for every $1\leq i+1<j+1$, so that $\mathcal{S}, w\models \langle\langle C\rangle\rangle \phi\mathsf{U}\psi$.

     \fussy
\end{itemize}

\end{proof}

To embed $ATL$ into stit theory, we present a proof system for $ATL$:

\begin{definition}[Proof system for $ATL$]
\label{axiomsatl}
Let $\Lambda_{\mathtt{ATL}}$ be the proof system defined by the following axioms and rules of inference---as presented by \citet{goranko2006complete}: 
\begin{itemize}
\item \emph{(Axioms)} All classical tautologies from propositional logic. For $A, B\subseteq Ags$ such that $A\cap B=\emptyset$, the following axiom schemata:
\[\begin{array}{ll}
\lnot \langle\langle A\rangle \rangle X\bot&(\bot)\\

\langle\langle A\rangle \rangle X\top&(\top)\\

\lnot \langle\langle \emptyset\rangle \rangle X\lnot p\to \langle\langle Ags\rangle \rangle X p&(GC)\\

\langle\langle A\rangle \rangle X p \land \langle\langle B\rangle \rangle X q \to \langle\langle A\cup B\rangle \rangle X( p\land  q  )&(S)\\

\langle\langle A\rangle \rangle \mathsf{G} p\leftrightarrow p\land \langle\langle A\rangle \rangle X\langle\langle A\rangle \rangle \mathsf{G} p  &(FP_\mathsf{G})\\

\langle\langle \emptyset\rangle \rangle \mathsf{G}( r\to( p\land\langle\langle A\rangle\rangle X r))\to \langle\langle \emptyset\rangle \rangle \mathsf{G}( r \to( p\land\langle\langle A\rangle\rangle\mathsf{G} p)) &(GFP_{\mathsf{G}})\\

\langle\langle A\rangle \rangle p\mathsf{U} q\leftrightarrow q\lor ( p\land\langle\langle A\rangle \rangle X\langle\langle A\rangle \rangle p \mathsf{U} q)  &(FP_\mathsf{U})\\ 

\langle\langle \emptyset\rangle \rangle \mathsf{G}(( q\lor( p\land \langle\langle A \rangle \rangle X r))\to  r)\to \langle\langle \emptyset\rangle \rangle \mathsf{G}(\langle\langle A \rangle \rangle  p \mathsf{U} q\to  r)\
&(LFP_{\mathsf{U}})\\ 

\end{array}\]
\item \textit{(Rules of inference)} \begin{itemize}
    \item \emph{Modus Ponens}, Substitution.
    \item \emph{$\langle\langle A\rangle\rangle X$-monotonicity}: from $ \phi\to  \psi$ infer $\langle\langle A\rangle\rangle X\phi\to \langle\langle A\rangle\rangle X\psi$.
    \item \emph{$\langle\langle \emptyset\rangle\rangle \mathsf{G}$-necessitation}: from $ \phi$ infer $\langle\langle \emptyset\rangle\rangle\mathsf{G} \phi$.
\end{itemize} 
\end{itemize}
\end{definition}

\begin{lemma}
\label{chingate}
 For each axiom and axiom schema $\phi$ of $\Lambda_{\mathtt{ATL}}$, $Tr(\phi)$ is a \emph{valid} formula with respect to the class of labelled \emph{bdt}-models.
\end{lemma}

\begin{proof}
Let $\mathcal{M}$ be a labelled \emph{bdt}-model.
\begin{itemize}
\sloppy
    \item As for the propositional tautologies, axiom schema $(\bot)$, and axiom schema $(\top)$, the property is immediate. 
    \item Let $\phi$ stand for axiom schema $(GC)$. Observe that $Tr(\phi)= \lnot \langle\langle \emptyset\rangle \rangle^s X \square p\to \langle\langle Ags\rangle \rangle^s X \square p$. This formula is equivalent to $\Diamond X \square p\to \langle\langle Ags\rangle \rangle^s X \square p$. Assume that $\mathcal{M},\langle m,h \rangle\models \Diamond X \square p$. This implies that there exists $h'\in H_{m}$ such that $\mathcal{M},\langle m,h' \rangle\models X\square  p$. Let $\sigma_{Ags}$ be a collective strategy starting at $m$ defined as follows: set $\sigma_{Ags}(m)=\mathbf{Choice}_{Ags}^m(h')$; for each $h''\in \mathbf{Choice}_{Ags}^m(h')$ and each $m'\in h''$ such that $m'\sqsupset m$, set $\sigma_{Ags}(m')$ as the choice-cell of $\mathbf{Choice}_{Ags}^{m'}$ including $h''$; for every \emph{other} $m'$ such that $m'\sqsupset m$, set $\sigma_{Ags}(m')$ as any choice-cell of $\mathbf{Choice}_{Ags}^{m'}$. It is clear that $\mathbf{Adm}^m_{Ags}\left({\sigma_{Ags}}\right)=\mathbf{Choice}_{Ags}^m(h')$, so that the fact that $\mathcal{M}$ is deterministic implies that $\mathcal{M},\langle m,h \rangle\models \langle\langle Ags\rangle \rangle^s X\square p$.
    
    \item  Let $\phi$ stand for axiom schema $(S)$. Then $Tr(\phi)= \langle\langle A\rangle \rangle^s X \square p \land \langle\langle B\rangle \rangle^s X \square q \to \langle\langle A\cup B\rangle \rangle^s X( \square p\land  \square q  )$. Assume that $\mathcal{M},\langle m,h \rangle\models  \langle\langle A\rangle \rangle^s X \square p \land \langle\langle B\rangle \rangle^s X\square q $. This implies the existence of collective strategies $\sigma_A$ and $\sigma_B$ starting at $m$ such that, for $h'\in  \mathbf{Adm}^m_{A}\left({\sigma_A}\right)$, $\mathcal{M},\langle m,h' \rangle\models X \square p$ and,  for $h''\in  \mathbf{Adm}^m_{B}\left({\sigma_B}\right)$, $\mathcal{M},\langle m,h'' \rangle\models X \square q$. Observe that the frame condition \emph{independence of agency} implies that $\sigma_A(m)\cap \sigma_B(m)$ is non-empty. Let $h_*\in \sigma_A(m)\cap \sigma_B(m)$. Let $\sigma_{A\cup B}$ be a collective strategy starting at $m$ defined as follows: set $\sigma_{A\cup B}(m)=\mathbf{Choice}_{A\cup B}^m(h_*)$; for each $h''\in \mathbf{Choice}_{A\cup B}^m(h_*)$ and each $m'\in h''$ such that $m'\sqsupset m$, set $\sigma_{A\cup B}(m')$ as the choice-cell of $\mathbf{Choice}_{A\cup B}^{m'}$ including $h''$; for every \emph{other} $m'$ such that $m'\sqsupset m$, set $\sigma_A\cup B(m')$ as any choice-cell of $\mathbf{Choice}_{A\cup B}^{m'}$. Let $h'\in\mathbf{Adm}^m_{A\cup B}\left({\sigma_{A\cup B}}\right)$. It is clear that $h'\in \mathbf{Choice}_{A\cup B}^m(h_*)$ so that $\mathcal{M},\langle m,h'\rangle \models X \square p\land X \square q$ and thus $\mathcal{M},\langle m,h \rangle\models \langle\langle A\cup B\rangle \rangle^s X (\square p\land \square q)$.
    
    \item   Let $\phi$ stand for axiom schema $(FP_{\mathsf{G}})$. Then $Tr(\phi)=\langle\langle A\rangle \rangle^s \mathsf{G} \square p\leftrightarrow \square p\land \langle\langle A\rangle \rangle^s X\langle\langle A\rangle \rangle^s \mathsf{G} \square p$. 
    
    For the right-to-left implication, assume that $\mathcal{M},\langle m,h \rangle\models \langle\langle A\rangle \rangle^s \mathsf{G} \square p$ via collective strategy $\sigma_A$ starting at $m$. It is straightforward to see that $\mathcal{M},\langle m,h \rangle\models  \square p$. Let $h'\in\mathbf{Adm}^m_{A}\left({\sigma_{A}}\right)$. Let $\sigma_{A}'$ be a collective strategy starting at $m^{+h'}$ defined as follows: for  $m'$ such that $m'\sqsupseteq m^{+h'}$, set $\sigma_A'(m')=\sigma_A(m')$. Let $h''\in\mathbf{Adm}^{m^{+h'}}_{A}\left(\sigma_A'\right)$. Observe that $h''\in H_{m^{+h'}}$ implies that $h''\in H_m$ and that $h''\in \mathbf{Choice}_{Ags}^m(h')$ (see Footnote \ref{bonham}). Furthermore, the definition of $\sigma_A'$ implies that $h''\in \sigma_A\left(m^{+h''(i)}\right)$ for every $i\in \mathds{N}$, which in turn implies that $h''\in\mathbf{Adm}^m_{A}\left({\sigma_{A}}\right)$. Our assumption then yields that $\mathcal{M},\langle m,h'' \rangle\models \mathsf{G} \square p$, so that $\mathcal{M},\langle m^{+h'},h'' \rangle\models \mathsf{G} \square p$ as well. Since $h''$ is an arbitrary history in $\mathbf{Adm}^{m^{+h'}}_{A}\left(\sigma_A'\right)$, this implies that $\mathcal{M},\langle m^{+h'},h' \rangle\models \langle\langle A\rangle\rangle^s \mathsf{G} \square p$, so that $\mathcal{M},\langle m,h' \rangle\models X \langle\langle A\rangle\rangle^s \mathsf{G} \square p$. Since $h'$ is an arbitrary history in $\mathbf{Adm}^m_{A}\left({\sigma_{A}}\right)$, this implies that $\mathcal{M},\langle m,h \rangle\models \langle\langle A\rangle\rangle^s X \langle\langle A\rangle\rangle^s \mathsf{G} \square p$. 
    
    For the left-to-right implication, assume that $\mathcal{M},\langle m,h \rangle\models  \square p\land \langle\langle A\rangle \rangle^s X\langle\langle A\rangle \rangle^s \mathsf{G} \square p$ via collective strategy $\sigma_A$ starting at $m$. This means that, for each $h'\in\mathbf{Adm}^m_{A}\left({\sigma_{A}}\right)$, $\mathcal{M},\langle m^{+h'},h' \rangle\models  \langle\langle A\rangle \rangle^s \mathsf{G} \square p$, where we take it that such a satisfaction is testified by collective strategy $\sigma_A^{h'}$ starting at $ m^{+h'}$.  Let $\sigma_{A}''$ be a collective strategy starting at $m$ defined as follows: set $\sigma_A''(m)=\sigma_A(m)$; for  $h_*\in \sigma_A(m)$ and $i\in \mathds{N}-\{0\}$, set $\sigma_A''\left(m^{+h_*(i)}\right)=\sigma_A^{h_*'}\left(m^{+h_*(i)}\right)$ for any $h_*'$ such that $h_*'\in \mathbf{Adm}^{m}_{A}\left({\sigma_{A}}\right)$.\footnote{The existence of such a history $h_*'$ is guaranteed by the fact that strategies map into cells of the choice partitions.} For $h_*\notin\sigma_A(m)$, set $\sigma_A''\left(m^{+h_*(i)}\right)=\sigma_A\left(m^{+h_*(i)}\right)$.

    Let $h''\in\mathbf{Adm}^m_{A}\left({\sigma_{A}''}\right)$. As implied by our assumption, it is the case that $\mathcal{M},\langle m^{+h''(0)},h'' \rangle\models \square p$. For $i\in\mathds{N}-\{0\}$, it is the case that $h''\in \sigma_A''\left(m^{+h''(i)}\right)= \sigma_A^{h_*'}\left(m^{+h''(i)}\right)$ (where $h_*'$ is the history such that $h_*'\in \mathbf{Adm}^m_{A}\left({\sigma_{A}}\right)$, $m^{+h_*'}=m^{+h''}$, and $\sigma_A^{h_*'}$ is the collective strategy starting at $m^{+h''}$ that testifies to the fact that $\mathcal{M},\langle m^{+h''},h_*' \rangle\models  \langle\langle A\rangle \rangle^s \mathsf{G} \square p$). This implies that $h''\in\mathbf{Adm}^{m^{+h''}}_{A}\left(\sigma_{A}^{h'_*}\right)$. Therefore, $\mathcal{M},\langle m^{+h''},h'' \rangle\models \mathsf{G}\square p$, thus supporting that $\mathcal{M},\langle m,h \rangle\models \langle\langle A\rangle \rangle^s \mathsf{G} \square p$.
    
    \item Let $\phi$ stand for axiom schema $(GFP_{\mathsf{G}})$. Then $Tr(\phi)=\langle\langle \emptyset\rangle \rangle^s \mathsf{G}( \square r\to( \square \square p\land\langle\langle A\rangle\rangle^s X  \square r))\to \langle\langle \emptyset\rangle \rangle^s \mathsf{G}( \square r \to( \square p\land\langle\langle A\rangle\rangle^s\mathsf{G} p))$. This formula is equivalent to $\square \mathsf{G}( \square r\to( \square p\land\langle\langle A\rangle\rangle^s X \square r))\to \square \mathsf{G}( \square r \to( \square p\land\langle\langle A\rangle\rangle^s\mathsf{G} \square p))$. Assume that $\mathcal{M},\langle m,h \rangle\models \square \mathsf{G}( \square r\to(\square p\land\langle\langle A\rangle\rangle^s X   \square r))$. 
    
    Let $h'\in H_m$ and $i\in \mathds{N}$. We want to show that $\mathcal{M},\left\langle m^{+h'(i)},h' \right\rangle\models \square r \to(\square p\land\langle\langle A\rangle\rangle^s\mathsf{G} \square p)$. Therefore, assume that $\mathcal{M},\left\langle m^{+h'(i)},h' \right\rangle\models \square r$. Our main assumption entails that $\mathcal{M},\left\langle m^{+h'(i)},h' \right\rangle\models  \square p\land\langle\langle A\rangle\rangle^s X \square r$. Let $\sigma_A$ be the collective strategy starting at $m^{+h'(i)}$ that testifies the satisfaction of the right conjunct. This means that, for each $h''\in \mathbf{Adm}^{m^{+h'(i)}}_{A}\left(\sigma_{A}\right)$, $\mathcal{M},\left\langle m^{+h'(i)},h'' \right\rangle\models X \square r$, which by the main assumption implies that $\mathcal{M},\left\langle m^{+h'(i)},h'' \right\rangle\models X (\square p\land \langle\langle A\rangle\rangle ^s X\square r)$, where $\sigma_A^{h''}$ will denote the collective strategy, starting at $\left(m^{+h'(i)}\right)^{+h''}$, that ensures the satisfaction of the formula ``$\langle\langle A\rangle\rangle ^s X\square r$'' at the successor moment of $m^{+h'(i)}$ along $h''$. Then, for each $h'''\in  \mathbf{Adm}^{\left(m^{+h'(i)}\right)^{+h''}}_{A}\left(\sigma_{A}^{h''}\right)$, one can apply the same argument and obtain a collective strategy $\sigma_A^{h'''}$ starting at $\left(\left(m^{+h'(i)}\right)^{+h''}\right)^{+h'''}$ that testifies to the fact that $\mathcal{M},\left\langle \left(m^{+h'(i)}\right)^{+h''},h''' \right\rangle\models X (\square p\land \langle\langle A\rangle\rangle ^s X\square r)$.

    Therefore, observe that for every  $m' \in\left\{m'; \mbox{ there is } h''\in \mathbf{Adm}^{m^{+h'(i)}}_{A}\left(\sigma_{A}\right) \mbox{ s.t. } m'=\left(m^{+h'(i)}\right)^{+h''}\right\}$, there exists $\sigma_A^{m'}$ such that for all the admissible histories of $\sigma_A^{m'}$, the successors of $m'$ along those admissible histories also satisfy the formula $\square p\land \langle\langle A\rangle\rangle ^s X\square r$, so that for each one of these successors there exists a collective strategy that does the same for its admissible histories, and so on and so forth.  
    
  Let us define a collective strategy $\sigma_A'$ starting at $m^{+h'(i)}$ as follows: set $\sigma_A'\left(m^{+h'(i)}\right)=\sigma_A \left(m^{+h'(i)}\right)$; for each $m'$ such that $m'\sqsupset m$, there are two cases: if $m'$ is a successor along an admissible history that does what was mentioned in the above paragraph, set $\sigma_A'(m')=\sigma_A^{m'}(m')$; if $m'$ is not any successor, then set $\sigma_A'(m')=\sigma_A(m')$. For $h''\in \mathbf{Adm}^{m^{+h'(i)}}_{A}\left(\sigma_{A}'\right) $, one can show that $\mathcal{M},\left\langle m^{+h'(i)},h'' \right\rangle\models \mathsf{G}\square p$. Therefore, we have shown that $\mathcal{M},\left\langle m^{+h'(i)},h' \right\rangle\models \square r \to(\square p\land\langle\langle A\rangle\rangle^s\mathsf{G} \square p)$, and thus that $\mathcal{M},\langle m,h \rangle\models \square \mathsf{G}( \square r \to( \square p\land\langle\langle A\rangle\rangle^s\mathsf{G} \square p))$.
  

  \item  Let $\phi$ stand for axiom schema $(FP_{\mathsf{U}})$. Then $Tr(\phi)=\langle\langle A\rangle \rangle^s \square  p\mathsf{U} \square q\leftrightarrow \square q\lor ( \square p\land\langle\langle A\rangle \rangle^s X\langle\langle A\rangle \rangle^s \square p \mathsf{U} \square q)$. 
  
  For the right-to-left implication, assume that $\mathcal{M},\langle m,h \rangle\models \langle\langle A\rangle \rangle^s \square p \mathsf{U}\square q $ via collective strategy $\sigma_A$ starting at $m$. Assume further that $\mathcal{M},\langle m,h \rangle\models \lnot\square q$. It is straightforward to see that our main assumption then implies that $\mathcal{M},\langle m,h \rangle\models  \square p$. Let $h'\in\mathbf{Adm}^m_{A}\left({\sigma_{A}}\right)$. Let $\sigma_{A}'$ be a collective strategy starting at $m^{+h'}$ defined as follows: for  $m'$ such that $m'\sqsupseteq m^{+h'}$, set $\sigma_A'(m')=\sigma_A(m')$. Let $h''\in\mathbf{Adm}^{m^{+h'}}_{A}\left(\sigma_A'\right)$. Observe that $h''\in H_{m^{+h'}}$ implies that $h''\in H_m$ and that $h''\in \mathbf{Choice}_{Ags}^m(h')$ (see Footnote \ref{bonham}). Furthermore, the definition of $\sigma_A'$ implies that $h''\in \sigma_A\left(m^{+h''(i)}\right)$ for every $i\in \mathds{N}$, which in turn implies that $h''\in\mathbf{Adm}^m_{A}\left({\sigma_{A}}\right)$. Our assumption then yields that $\mathcal{M},\langle m,h'' \rangle\models \square p \mathsf{U}\square q $. Since it is also that case that $\mathcal{M},\langle m,h'' \rangle\models \lnot\square q$ (as implied by the ``further'' assumption made above), then one gets that $\mathcal{M},\langle m^{+h'},h'' \rangle\models \square p \mathsf{U}\square q$ as well. Since $h''$ is an arbitrary history in $\mathbf{Adm}^{m^{+h'}}_{A}\left(\sigma_A'\right)$, this implies that $\mathcal{M},\langle m^{+h'},h' \rangle\models \langle\langle A\rangle\rangle^s  \square p \mathsf{U}\square q$, so that $\mathcal{M},\langle m,h' \rangle\models X \langle\langle A\rangle\rangle^s \square p \mathsf{U}\square q$. Since $h'$ is an arbitrary history in $\mathbf{Adm}^m_{A}\left({\sigma_{A}}\right)$, this implies that $\mathcal{M},\langle m,h \rangle\models \langle\langle A\rangle\rangle^s X \langle\langle A\rangle\rangle^s  \square p \mathsf{U}\square q$. 
  
  For the left-to-right implication, assume that $\mathcal{M},\langle m,h \rangle\models  \square q\lor ( \square p\land\langle\langle A\rangle \rangle^s X\langle\langle A\rangle \rangle^s \square p \mathsf{U} \square q)$. There are two cases, then: 
  
  \begin{itemize}
      \item If $\mathcal{M},\langle m,h \rangle\models  \square q$, it is clear that $\mathcal{M},\langle m,h \rangle\models  \square\square q$, which turn implies that $\mathcal{M},\langle m,h \rangle\models  \square(\square p\mathsf{U}\square q)$, which in turn implies that $\mathcal{M},\langle m,h \rangle\models  \langle\langle A\rangle \rangle^s\square p \mathsf{U}\square q$.
      
      \item Assume that $\mathcal{M},\langle m,h \rangle\models \square p\land\langle\langle A\rangle \rangle^s X\langle\langle A\rangle \rangle^s \square p \mathsf{U} \square q$, so that the satisfaction of the right conjunct is testified by collective strategy $\sigma_A$ starting at $m$.  This means that, for each $h'\in\mathbf{Adm}^m_{A}\left({\sigma_{A}}\right)$, $\mathcal{M},\langle m^{+h'},h' \rangle\models  \langle\langle A\rangle \rangle^s \square p \mathsf{U} \square q$, where we take it that such a satisfaction is testified by collective strategy $\sigma_A^{h'}$ starting at $ m^{+h'}$. Let $\sigma_{A}''$ be a collective strategy starting at $m$ defined as follows: set $\sigma_A''(m)=\sigma_A(m)$; for $h_*\in \sigma_A(m)$ and $i\in \mathds{N}-\{0\}$, set $\sigma_A''\left(m^{+h_*(i)}\right)=\sigma_A^{h_*'}\left(m^{+h_*(i)}\right)$; if $h_*\not\in \mathbf{Adm}^m_{A}\left({\sigma_{A}}\right)$, then set $\sigma_A''\left(m^{+h_*(i)}\right)=\sigma_A\left(m^{+h_*(i)}\right)$. 
      
      Let $h''\in\mathbf{Adm}^m_{A}\left({\sigma_{A}''}\right)$. 
        As implied by our assumption, it is the case that ($\star$) $\mathcal{M},\langle m^{+h''(0)},h'' \rangle\models \square p$.  For $i\in\mathds{N}-\{0\}$, it is the case that $h''\in \sigma_A''\left(m^{+h''(i)}\right)= \sigma_A^{h_*'}\left(m^{+h''(i)}\right)$ (where $h_*'$ is the history such that $h_*'\in \mathbf{Adm}^m_{A}\left({\sigma_{A}}\right)$, $m^{+h_*'}=m^{+h''}$, and $\sigma_A^{h_*'}$ is the collective strategy starting at $m^{+h''}$ that testifies to the fact that $\mathcal{M},\langle m^{+h''},h_*' \rangle\models  \langle\langle A\rangle \rangle^s \mathsf{G} \square p$). This implies that $h''\in\mathbf{Adm}^{m^{+h''}}_{A}\left(\sigma_{A}^{h'_*}\right)$. Therefore, $\mathcal{M},\langle m^{+h''},h'' \rangle\models \square p\mathsf{U}\square q$, which  with $(\star)$  straightforwardly implies that $\mathcal{M},\langle m,h'' \rangle\models \square p\mathsf{U}\square q$, thus supporting that $\mathcal{M},\langle m,h \rangle\models \langle\langle A\rangle \rangle^s \square p \mathsf{U} \square q$.
        
        \item Let $\phi$ stand for axiom schema $(LFP_{\mathsf{U}})$. Then $Tr(\phi)=\langle\langle \emptyset\rangle \rangle^s \mathsf{G}(( \square q\lor( \square p\land \langle\langle A \rangle \rangle^s X \square r))\to  \square r)\to \langle\langle \emptyset\rangle \rangle^s \mathsf{G}(\langle\langle A \rangle \rangle^s \square p \mathsf{U} \square q\to  \square r)$. This formula is equivalent to $\square \mathsf{G}(( \square q\lor( \square p\land \langle\langle A \rangle \rangle^s X \square r))\to  \square r)\to \square \mathsf{G}(\langle\langle A \rangle \rangle^s \square p \mathsf{U} \square q\to  \square r)$. 
        
        Assume that $\mathcal{M},\langle m,h \rangle\models \square \mathsf{G}(( \square q\lor( \square p\land \langle\langle A \rangle \rangle^s X \square r))\to  \square r)$.  Let $h'\in H_m$ and $i\in \mathds{N}$. We want to show that $\mathcal{M},\left\langle m^{+h'(i)},h' \right\rangle\models \langle\langle A \rangle \rangle^s \square p \mathsf{U} \square q\to  \square r$. Therefore, assume further that $\mathcal{M},\left\langle m^{+h'(i)},h' \right\rangle\models  \langle\langle A \rangle \rangle^s \square p \mathsf{U} \square q$ via collective strategy $\sigma_A$.
        
        If there exists $h''\in \mathbf{Adm}^{m^{+h'(i)}}_{A}\left(\sigma_{A}\right)$ such that $\mathcal{M},\left\langle m^{+h'(i)},h'' \right\rangle\models \square q$, then the main assumption implies straightforwardly that $\mathcal{M},\left\langle m^{+h'(i)},h'' \right\rangle\models \square r$, yielding what we want. 
 
 Thus, assume that for each $h''\in \mathbf{Adm}^{m^{+h'(i)}}_{A}\left(\sigma_{A}\right)$, the index $j$ such that $\mathcal{M},\left\langle \left(m^{+h'(i)}\right)^{+h''(j)},h'' \right\rangle\models \square q$ is strictly greater than $0$. Let $h''\mathbf{Adm}^{m^{+h'(i)}}_{A}\left(\sigma_{A}\right)$ be such that the aforementioned index is $j$.  We will show that $\mathcal{M},\left\langle \left(m^{+h'(i)}\right),h'' \right\rangle\models \square p\land \langle\langle A \rangle \rangle^s X\square r$. The further assumption that $\mathcal{M},\left\langle m^{+h'(i)},h'' \right\rangle\models  \langle\langle A \rangle \rangle^s \square p \mathsf{U} \square q$ via collective strategy $\sigma_A$ implies that $\mathcal{M},\left\langle \left(m^{+h'(i)}\right),h'' \right\rangle\models \square p$ straightforwardly. 
 
 For the other conjunct, let us show that the following claim is true: for each $h'''\in  \mathbf{Adm}^{m^{+h'(i)}}_{A}\left(\sigma_{A}\right)$, $\mathcal{M},\left\langle \left(m^{+h'(i)}\right)^{+h'''},h''' \right\rangle\models \square r$. Assume for a contradiction that this is not the case. Then there exists an admissible history $h_*$ such that $\mathcal{M},\left\langle \left(m^{+h'(i)}\right)^{+h_*},h_* \right\rangle\not\models \square r$. By the main assumption, this implies that $\square q$ does not hold at such an index, so that the further assumption that $\mathcal{M},\left\langle m^{+h'(i)},h'' \right\rangle\models  \langle\langle A \rangle \rangle^s \square p \mathsf{U} \square q$ via collective strategy $\sigma_A$ implies that $\square p$ holds at such an index. Observe that it \emph{cannot} be the case that, for all $h'''\in \mathbf{Adm}^{m^{+h'(i)}}_{A}\left(\sigma_{A}\right)$ such that $\left(m^{+h'(i)}\right)^{+h_*}\in h'''$, $\square r$ holds at the index given by the successor of $\left(m^{+h'(i)}\right)^{+h_*}$ along $h'''$ and $h'''$, since this would imply that $\langle\langle A \rangle \rangle^s X\square r$ holds at every index based on $\left(m^{+h'(i)}\right)^{+h_*}$, and the main assumption would therefore imply that $\square r$ holds at $\left\langle \left(m^{+h'(i)}\right)^{+h_*},h_* \right\rangle$, something that we have assumed to not happen. Therefore, there must exist $h_*'\in \mathbf{Adm}^{m^{+h'(i)}}_{A}\left(\sigma_{A}\right)$ such that $\mathcal{M},\left\langle \left(\left(m^{+h'(i)}\right)^{+h_*}\right)^{+h_*'},h_*' \right\rangle\not\models \square r$. Again, the main assumption then implies that $\square q$ does not hold at such an index, so that the further assumption that $\mathcal{M},\left\langle m^{+h'(i)},h'' \right\rangle\models  \langle\langle A \rangle \rangle^s \square p \mathsf{U} \square q$ via collective strategy $\sigma_A$ implies that $\square p$ holds at such an index. Analogously, it \emph{cannot} be the case that, for all $h'''\in \mathbf{Adm}^{m^{+h'(i)}}_{A}\left(\sigma_{A}\right)$ such that $\left(\left(m^{+h'(i)}\right)^{+h_*}\right)^{+h_*'}\in h'''$, $\square r$ holds at the index given by the successor of $\left(\left(m^{+h'(i)}\right)^{+h_*}\right)^{+h_*'}$ along $h'''$ and $h'''$, so there must exist an admissible history $h_*''$ such that $\square r$ does not hold at the index given by the successor of $\left(\left(m^{+h'(i)}\right)^{+h_*}\right)^{+h_*'}$ along $h_*''$ and $h_*''$. Again, the main assumption then implies that $\square q$ does not hold at such an index, so that the further assumption that $\mathcal{M},\left\langle m^{+h'(i)},h'' \right\rangle\models  \langle\langle A \rangle \rangle^s \square p \mathsf{U} \square q$ via collective strategy $\sigma_A$ implies that $\square p$ holds at such an index. With this argument, and natural induction, one can show that there exists a history $h_+\in  \mathbf{Adm}^{m^{+h'(i)}}_{A}\left(\sigma_{A}\right)$ such that $\mathcal{M},\left\langle m^{+h'(i)},h_+ \right\rangle\models  \mathsf{G} \square p$, contradicting the assumption that $\sigma_A$ testifies to the fact that $\mathcal{M},\left\langle m^{+h'(i)},h'' \right\rangle\models  \langle\langle A \rangle \rangle^s \square p \mathsf{U} \square q$. Therefore, $\mathcal{M},\left\langle m^{+h'(i)},h'' \right\rangle\models  \langle\langle A \rangle \rangle^s X\square r$.
 
 Therefore, it is the case that $\mathcal{M},\left\langle m^{+h'(i)},h'' \right\rangle\models  \square p \land\langle\langle A \rangle \rangle^s X\square r$. The main assumption then implies that $\mathcal{M},\left\langle m^{+h'(i)},h'' \right\rangle\models  \square r$, so that $\mathcal{M},\left\langle m^{+h'(i)},h' \right\rangle\models  \square r$ as well.

  \end{itemize}

\end{itemize}
\fussy
\end{proof}

\begin{lemma}\label{vetealaverga}
Let formula $\phi, \psi$ be formulas of the language $\mathcal{L}_{\textsf{ATL}}$, then the following clauses hold:
\begin{enumerate}[a)]
    \item If $Tr(\phi)$ and $Tr(\phi\to \psi)$ are valid with respect to the class of labelled \emph{bdt}-models, then $Tr(\psi)$ is also valid with respect to the class of labelled \emph{bdt}-models.
    
    \item If $\psi$ is obtained from $\phi$ by a uniform substitution of the propositional letters in $\phi$ with arbitrary formulas of $\mathcal{L}_{\textsf{ATL}}$, and if $Tr(\phi)$ is valid with respect to the class of labelled \emph{bdt}-models, then  $Tr(\psi)$ is also valid with respect to the class of labelled \emph{bdt}-models.
    
    \item If $Tr(\phi\to \psi)$ is valid with respect to the class of labelled \emph{bdt}-models, then 
    
    $Tr\left(\langle\langle A\rangle\rangle X\phi\to \langle\langle A\rangle\rangle X\psi\right)$ is also valid with respect to the class of labelled \emph{bdt}-models.

    \item If $Tr(\phi)$ is valid with respect to the class of labelled \emph{bdt}-models, $Tr(\langle\langle\emptyset\rangle\rangle\mathsf{G}\phi)$ is also valid with respect to the class of labelled \emph{bdt}-models.
\end{enumerate}
\end{lemma}
\begin{proof}

\begin{enumerate}[a)]

\item Observe that $Tr(\phi\to \psi)$ is logically equivalent to $Tr(\phi)\to Tr(\psi)$. Therefore if $Tr(\phi)$ is valid with respect to the class of labelled \emph{bdt}-models, and $Tr(\phi)\to Tr(\psi)$ as well, we get that $Tr(\psi)$ is also valid with respect to the class of labelled \emph{bdt}-models.

\item Let $p_i\mapsto \theta_i$ be the uniform substitution on $\phi$. Observe that $Tr(\psi)$ is obtained by the uniform substitution $p_i\mapsto Tr(\theta_i)$. Since uniform substitution preserves validity with respect to the class of labelled \emph{bdt}-models, the fact that $Tr(\phi)$ is valid with respect to the class of labelled \emph{bdt}-models implies that $Tr(\psi)$ is also valid with respect to the class of labelled \emph{bdt}-models.

\item Follows straightforwardly from the facts that $Tr(\phi\to \psi)$ is logically equivalent to $Tr(\phi)\to Tr(\psi)$ and that  $Tr\left(\langle\langle A\rangle\rangle X\phi\to \langle\langle A\rangle\rangle X\psi\right)$ is logically equivalent to $\langle\langle A\rangle\rangle^s XTr(\phi)\to \langle\langle A\rangle\rangle^s XTr(\psi)$. 

\item Follows straightforwardly from the fact that $Tr(\langle\langle\emptyset\rangle\rangle\mathsf{G}\phi)$ is logically equivalent to $\square \mathsf{G}Tr(\phi)$. 

\end{enumerate}
\end{proof}

\begin{proposition} \label{chingadamadre}
 For a formula $\phi$ of the language $\mathcal{L}_{\textsf{ATL}}$, it is the case that that $\vdash_{\Lambda_{\mathtt{ATL}}}\phi$ iff $Tr(\phi)$ is \emph{valid} with respect to the class of deterministic labelled \emph{bdt}-models. 
\end{proposition}
\begin{proof}
($\Rightarrow$) Assume that $\vdash_{\Lambda_{\mathtt{ATL}}}\phi$. Let us proceed by induction on the length of the proof. Lemma \ref{chingate} ensures that if $\phi$ is an instance of an axiom or axiom schema of $\Lambda_{\mathtt{ATL}}$, then $Tr(\phi)$ is valid with respect to the class of labelled \emph{bdt}-models. If $\phi$ is obtained from the theorems of $\Lambda_{\mathtt{ATL}}$ by uniform substitution, \emph{Modus Ponens}, \emph{$\langle\langle A\rangle\rangle X$-monotonicity}, or \emph{$\langle\langle \emptyset\rangle\rangle \mathsf{G}$-necessitation}, then Lemma \ref{vetealaverga} ensures that $Tr(\phi)$ is valid with respect to the class of labelled \emph{bdt}-models.

($\Leftarrow$) We work by contraposition. Assume that $\nvdash_{\Lambda_{\mathtt{ATL}}}\phi$. This means that $\lnot \phi$ is $\Lambda_{\mathtt{ATL}}$-consistent. By \citeauthor{goranko2006complete}'s \citeyearpar{goranko2006complete}  result of completeness, there exists a concurrent game structure $\mathcal{S}$ such that $\mathcal{S},w\models \lnot \phi$ for some $w$ in the domain of  $\mathcal{S}$. By Proposition \ref{gottabe}, $\mathcal{M}^{\mathcal{S}}, \langle\lambda_w, h \rangle\models Tr(\lnot\phi)$ for every $h\in H_{\lambda_w}$. Since one such $h$ exists by construction (let us denote it by $h_*$), it is the case that $\mathcal{M}^{\mathcal{S}}, \langle\lambda_w, h_* \rangle\models \lnot Tr(\phi)$, so that $Tr(\phi)$ is not valid with respect to the class of  labelled \emph{bdt}-models. 

\end{proof}

\begin{restatable}{proposition}{valatl} \label{madres}
For a formula $\phi$ of the language $\mathcal{L}_{\textsf{ATL}}$, $\phi$ is \emph{valid} with respect to the class of concurrent game structures iff $Tr(\phi)$ is \emph{valid} with respect to the class of deterministic labelled \emph{bdt}-models. 
\end{restatable}
\begin{proof}

($\Rightarrow$) If $\phi$ is \emph{valid} with respect to the class of concurrent game structures, then  \citeauthor{goranko2006complete}'s \citeyearpar{goranko2006complete} result of completeness implies  that $\vdash_{\Lambda_{\mathtt{ATL}}}\phi$. The left-to-right direction of Proposition \ref{chingadamadre} then implies that $Tr(\phi)$ is valid with respect to the class of deterministic labelled \emph{bdt}-models.

($\Leftarrow$) If $Tr(\phi)$ is \emph{valid} with respect to the class of deterministic labelled s\emph{bdt}-models, then by the right-to-left direction of Proposition \ref{chingadamadre} it is the case that $\vdash_{\Lambda_{\mathtt{ATL}}}\phi$. \citeauthor{goranko2006complete}'s \citeyearpar{goranko2006complete} result of soundness then implies that $\phi$ is valid with respect to the class of concurrent game structures. 

\end{proof}

\bibliographystyle{apalike}

\bibliography{references}

\end{document}